\documentclass[PAPER, atlasdraft=false, texlive=2025, UKenglish, cernpreprint, texmf, orcidlogo]{atlasdoc}%
\usepackage{atlaspackage}
\usepackage{atlasbiblatex}

\usepackage{atlasphysics}
\graphicspath{{logos/}{figures/}}

\usepackage{atlascover}
\AtlasTitle{Search for emerging jets in $pp$ collisions at $\sqrt{s} = 13~\mathrm{TeV}$ with the ATLAS experiment}

\AtlasAbstract{%
A search is presented for emerging jets using $140~\mathrm{fb}^{-1}$ of proton--proton collision data at $\sqrt{s} = 13~\mathrm{TeV}$, collected by the ATLAS experiment between 2015 and 2018. The search looks for the existence of a dark sector with symmetries similar to those in quantum chromodynamics. This dark sector is populated with dark quarks, which undergo showering similar to quarks in the Standard Model, leading to a high multiplicity of long-lived dark hadrons within a dark jet. These dark hadrons subsequently decay to Standard Model particles via a new heavy scalar mediating particle $\phi$. This results in jets which contain multiple displaced vertices, known as emerging jets. This analysis targets four-jet topologies, with two emerging jets and two Standard Model jets, resulting from the decay of pair-produced scalar mediators. No significant excess above the Standard Model background is observed. For dark pion proper decay lengths of $20~\mathrm{mm}$, mediator masses are excluded between $1$ and $2~\mathrm{TeV}$ assuming a dark pion mass of $20~\mathrm{GeV}$.
}

\AtlasRefCode{EXOT-2018-20}

\PreprintIdNumber{CERN-EP-2025-225}

\arXivId{2510.12347}

\HepDataRecord{https://www.hepdata.net/record/ins3069355?version=1}

\AtlasJournalRef{Eur. Phys. J. C 86 (2026) 808}
\AtlasDOI{10.1140/epjc/s10052-026-15693-1}

\AtlasCoverEgroupAnalysisTeam{atlas-EXOT-2018-20-analysis-team@cern.ch}

\hypersetup{pdftitle={ATLAS document},pdfauthor={The ATLAS Collaboration}}

\begin{document}

\maketitle
\setcounter{tocdepth}{1}
\tableofcontents

\section{Introduction}

Although there is a vast body of evidence for the existence of Dark Matter (DM)~\cite{Rubin_1980,Persic_1996,Clowe_2006}, its nature is one of the major unanswered questions in fundamental physics. Many theoretical models of particle DM have been proposed, each with corresponding interactions with the Standard Model (SM). However, after years of searching, no experiment has yet provided evidence of the particle nature of DM~\cite{DarkSide,SuperCDMS,PICO,XENON,DEAP-3600,PandaX,EXOT-2023-14,CMS-EXO-23-005,LZ}. This drought of experimental confirmations has prompted the search for more exotic models of DM, including the existence of a \textit{dark sector} (DS). The DS is an extension of the SM which contains new particles that are uncharged under the symmetries of the SM, instead having interactions governed by a new confining DS gauge group~\cite{Strassler_2007,Strassler_2008,Han_2008}. Although the DS and SM cannot communicate directly, interactions between DS and SM particles can be mediated by new particles that are charged under both SM and DS symmetries.

Several models propose new DS interactions which resemble quantum chromodynamics (QCD) in the SM, colloquially known as \textit{dark QCD}~\cite{Bai_2014,Beauchesne_2019,Albouy_2022}. These DS models contain quark-like particles, with $N_{f}$ flavours and $N_{c}$ colours, which transform under a non-Abelian gauge group SU($N_{c}$). This new interaction is characterized by a confining scale ($\Lambda_{d}$) below which dark quarks ($Q_{d}$) hadronize, analogously to QCD, to form dark baryons and mesons. The lightest dark baryon could be stable, providing a DM candidate which conforms to the astrophysical evidence, while the lightest dark mesons, \textit{dark pions} ($\pi_{d}$), would be unstable and decay to into SM particles via an exchange of a virtual mediator particle. Depending on the lifetimes of the dark pions, the resulting decay topology can differ greatly, giving rise to three general categories of detector signatures in collider experiments: prompt \textit{dark jets}~\cite{darkJets}, \textit{semi-visible jets}~\cite{Cohen_2015}, and \textit{emerging jets} (EJs)~\cite{SSW}. In each case there is some component of the DS which is unstable, and therefore decays back to the SM, and so dark QCD signatures are characterized by a double shower and hadronization, once in the DS and once in the SM which gives rise to wider jets with higher charged-track multiplicities. For the case of dark jets, where all the dark hadrons decay promptly back to SM particles, these features allow them to be differentiated from SM hadronic jets. In the case of semi-visible jets, however, some of the dark hadrons decay promptly, while others are stable, and therefore carry missing transverse energy (\MET) which goes undetected; this results in \MET aligned with the hadronic jet which can be used to distinguish semi-visible jets from SM jets. Finally, in the case of EJs, all of the dark hadrons are unstable, but are also long-lived, and therefore they decay within the detector volume but displaced from the initial collision, giving rise to multiple displaced vertices (DVs) contained in the jet cone. Due to the exponential distribution of the $\pi_{d}$ decay times, these DVs occur at varying radial distances, giving rise to the emerging quality of the EJ.

Using the proton--proton ($pp$) collisions of the LHC, both ATLAS and CMS collaborations have carried out dedicated searches for different dark-QCD scenarios, characterized by various production modes and decay topologies. Both ATLAS and CMS have searched for semi-visible jets in Run 2 data~\cite{EXOT-2022-37,EXOT-2021-19,CMS-EXO-19-020}. ATLAS has set limits on a search for promptly decaying dark hadrons from the resonant production of dark jets~\cite{HDBS-2018-45}. In Run 1, ATLAS performed a search for hadronic decays of long-lived particles which give a displaced jet signature~\cite{EXOT-2013-12}. In Run 2, CMS performed two searches specifically targeting EJ signatures by looking for pair-produced, \textit{bifundamental}\footnote{Bifundamental means that it is charged under both the symmetries of Dark QCD and Standard Model QCD as $SU(3)_{QCD} = 3$ and $SU(N_{c})_{Dark} = N_{c}$. Note that the bifundamental scalar is labeled $X_{d}$ in Ref.~\cite{SSW} and $X_{\mathrm{dark}}$ in Ref.~\cite{CMS-EXO-18-001,CMS-EXO-22-015}.}, scalar mediators ($\phi$) resulting in a four-jet topology~\cite{CMS-EXO-18-001,CMS-EXO-22-015}. In Run 3 ATLAS produced a search for EJs that focused on the resonant production of a vector mediator ($Z'$) resulting in a two-jet topology~\cite{EXOT-2022-31}. The Run 3 analysis also included an interpretation using $t-$channel production of the bifundamental scalar mediator with two, three, or four jets in the final state, which causes the production cross-section to be dependent on the Yukawa coupling of the mediator. For dark pions with proper decay length of $10~\mathrm{mm}$ and mass of $10~\GeV$, the CMS search produced limits which excluded scalar mediator masses up to $1.95~\TeV$, and the ATLAS search produced limits which excluded vector mediator masses up to $2.55~\TeV$.

\begin{figure}[h]
\centering
\includegraphics[width=0.7\textwidth]{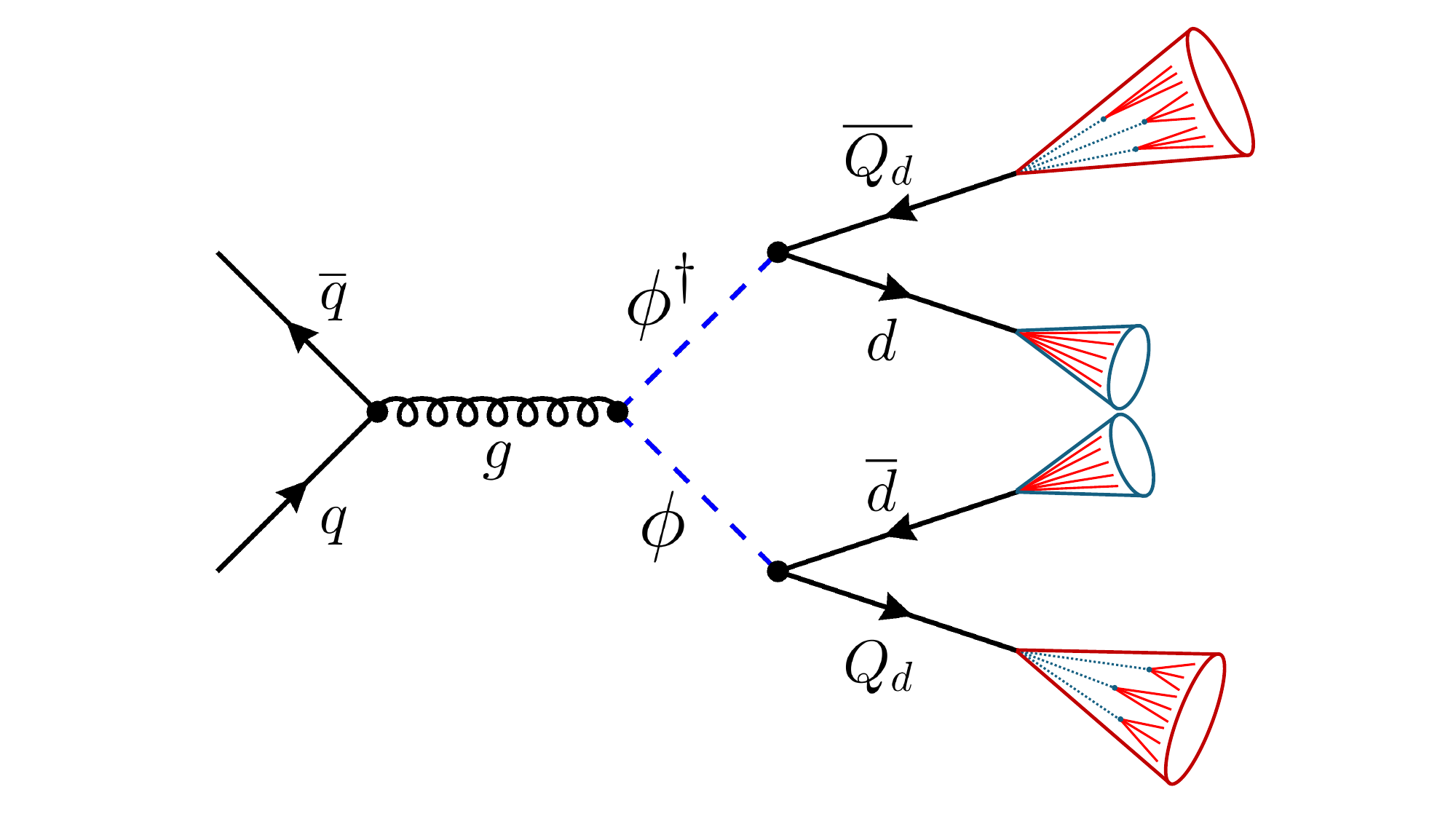}
\caption{Feynman diagram of the pair-production of dark scalar mediators, $\phi$, via quark--antiquark annihilation. The resulting four jets are shown as cones for illustrative purposes. Dashed lines within the cones represent the trajectories of dark pions, solid lines within the cones represent the trajectories of SM particles.}
\label{fig:EJfeynman}
\end{figure}

This search looks for pair-produced dark scalar mediators in $pp$ collisions at $\sqrt{s} = 13~\TeV$. The two production modes considered are $q\overline{q}$ annihilation (shown in Figure~\ref{fig:EJfeynman}) and gluon--gluon fusion. These production modes are expected to dominate over $t-$channel production used in the Run 3 search~\cite{SSW}. Both modes proceed via a virtual gluon which produces a pair of dark mediators ($\phi^{\dagger}\phi$). Each dark mediator immediately decays into a dark quark and a SM quark, both of which hadronize in their corresponding sectors, resulting in a SM jet and an EJ. The final topology is a four jet final state, with two emerging jets and two SM jets. The Yukawa coupling ($\kappa$) is assumed to be sufficiently small, that diagrams proportional to $\kappa^{4}$ are ignored.

An \textit{unflavoured} model (similar to the most-minimal model in Ref.~\cite{tChannel}) is considered, where the Yukawa coupling between the dark mediator and the SM quark is only non-zero for SM down quarks\footnote{In addition to $SU(3)_{QCD} = 3$ charge, the mediator has SM $U(1) = \frac{1}{3}$ charge, since it couples to down quarks.}, giving only light SM jets in the final state. The choice of $\phi$ coupling only to SM down quarks or SM up quarks is arbitrary in the unflavoured model, as both will give only light SM jets in the final state; the down-type coupling is chosen following Ref.~\cite{SSW}. The Run 3 search considered a \textit{flavour-aligned} model of the bifundamental scalar, where the scalar mediator can couple to any down-type standard model quark. This allows for heavy-flavour jets in the final state when kinematically allowed.

A two-stage machine learning (ML) based classifier is used to separate EJ-like events from the dominant QCD multijet background. The contribution from this background is estimated using a data-driven approach based on the ABCD method. Other background processes, such as top quark pair-production, were considered, but their contributions are suppressed to a negligible level by analysis preselections.

The paper is structured in the following way: Section~\ref{sec:detector} gives a description of the Run 2 configuration of the ATLAS detector, while Section~\ref{sec:dataMC} gives an overview of the collected dataset, and the simulated samples produced for this analysis. Sections~\ref{sec:reconstruction} and~\ref{sec:selection} detail how the various physics objects are reconstructed, and how events are selected for the analysis. The ML classifier is described in Section~\ref{sec:analysis}. Then, Section~\ref{sec:background} outlines the background estimation method and Section~\ref{sec:systematics} documents the systematic uncertainties. Finally, the results of the analysis are presented in Section~\ref{sec:result} with conclusions given in Section~\ref{sec:conclusion}.

\section{ATLAS detector}
\label{sec:detector}

\newcommand{\AtlasCoordFootnote}{%
ATLAS uses a right-handed coordinate system with its origin at the nominal interaction point (IP) in the centre of the detector and the \(z\)-axis along the beam pipe. The \(x\)-axis points from the IP to the centre of the LHC ring, and the \(y\)-axis points upwards. Polar coordinates \((r,\phi)\) are used in the transverse plane, \(\phi\) being the azimuthal angle around the \(z\)-axis. The pseudorapidity is defined in terms of the polar angle \(\theta\) as \(\eta = -\ln \tan(\theta/2)\) and is equal to the rapidity $ y = \frac{1}{2} \ln \left( \frac{E + p_z}{E - p_z} \right) $ in the relativistic limit. Angular distance is measured in units of \(\Delta R \equiv \sqrt{(\Delta y)^{2} + (\Delta\phi)^{2}}\).}

The ATLAS detector~\cite{PERF-2007-01} at the LHC covers nearly the entire solid angle around the collision point.\footnote{\AtlasCoordFootnote} It consists of an inner tracking detector surrounded by a thin superconducting solenoid, electromagnetic and hadronic calorimeters, and a muon spectrometer incorporating three large superconducting air-core toroidal magnets.

The inner-detector system (ID) is immersed in a \qty{2}{\tesla} axial magnetic field and provides charged-particle tracking in the range \(|\eta| < 2.5\). The high-granularity silicon pixel detector covers the vertex region and typically provides four measurements per track, the first hit generally being in the insertable B-layer (IBL) installed before Run~2~\cite{ATLAS-TDR-19,PIX-2018-001}. It is followed by the SemiConductor Tracker (SCT), which usually provides eight measurements per track. These silicon detectors are complemented by the transition radiation tracker (TRT),
which enables radially extended track reconstruction up to \(|\eta| = 2.0\).
The TRT also provides electron identification information
based on the fraction of hits (typically 30 in total) above a higher energy-deposit threshold corresponding to transition radiation.

The calorimeter system covers the pseudorapidity range \(|\eta| < 4.9\). Within the region \(|\eta|< 3.2\), electromagnetic calorimetry is provided by barrel and endcap high-granularity lead/liquid-argon (LAr) calorimeters,
with an additional thin LAr presampler covering \(|\eta| < 1.8\)
to correct for energy loss in material upstream of the calorimeters.
Hadronic calorimetry is provided by the steel/scintillator-tile calorimeter,
segmented into three barrel structures within \(|\eta| < 1.7\), and two copper/LAr hadronic endcap calorimeters. The solid angle coverage is completed with forward copper/LAr and tungsten/LAr calorimeter modules optimised for electromagnetic and hadronic energy measurements, respectively.

The muon spectrometer (MS) comprises separate trigger and
high-precision tracking chambers measuring the deflection of muons in a magnetic field generated by the superconducting air-core toroidal magnets.
The field integral of the toroids ranges between \num{2.0} and \qty{6.0}{\tesla\metre} across most of the detector. Three layers of precision chambers, each consisting of layers of monitored drift tubes, cover the region \(|\eta| < 2.7\), complemented by cathode-strip chambers in the forward region, where the background is highest. The muon trigger system covers the range \(|\eta| < 2.4\) with resistive-plate chambers in the barrel, and thin-gap chambers in the endcap regions.

The luminosity is measured mainly by the LUCID--2~\cite{LUCID2} detector that records Cherenkov light produced in the quartz windows of photomultipliers located close to the beampipe.

Events were selected by the first-level trigger system implemented in custom hardware, followed by selections made by algorithms implemented in software in the high-level trigger~\cite{TRIG-2016-01}. The first-level trigger accepted events from the \qty{40}{\MHz} bunch crossings at a rate close to \qty{100}{\kHz}, which the high-level trigger further reduced in order to record complete events to disk at about \qty{1.25}{\kHz}.

A software suite~\cite{SOFT-2010-01} is used in data simulation, in the reconstruction and analysis of real and simulated data, in detector operations, and in the trigger and data acquisition
systems of the experiment.

\section{Data and Simulated Events}
\label{sec:dataMC}

This analysis uses $140~\ifb$ of $pp$ collision data collected at $\sqrt{s}=13~\TeV$ by the ATLAS experiment between the years of 2015 and 2018, corresponding to Run 2 of the LHC. The dataset contains only those events recorded during stable beam conditions, with all detector subsystems operating properly. Standard data-quality requirements are used to ensure reliable reconstruction of the physics objects~\cite{DAPR-2018-01}.

Monte Carlo (MC) simulation events, generated with \PYTHIA[8.230]~\cite{Sjostrand:2014zea}, are used to optimise the analysis strategy, evaluate the signal efficiency, and test the background estimation methods. Matrix element generation of the hard scattering process and subsequent showering uses the \NNPDF[2.3lo]~\cite{Ball:2012cx} parton distribution function (PDF) set, with the A14 set of tuned parameters~\cite{ATL-PHYS-PUB-2014-021}. Pile-up, the effect of multiple interactions in the same and neighbouring bunch crossings, is modelled by combining the results of the simulated hard-scattering event with simulated inelastic $pp$ events generated with \PYTHIA[8.186]~\cite{Sjostrand:2007gs} and the A3 set of tuned parameters~\cite{ATL-PHYS-PUB-2016-017}. The interaction of the collision products with the ATLAS detector, and subsequent detector response~\cite{SOFT-2010-01}, are modelled by passing MC events through a detailed map of the detector material implemented into \textsc{Geant4}~\cite{Agostinelli:2002hh}. MC events then are reconstructed using the same algorithms applied to the data.

\subsection{Signal Simulation}

Samples of $pp \to \phi^{\dagger}\phi \to \overline{Q}_{d}qQ_{d}\overline{q}$ are produced at leading-order (LO) using the Hidden Valley~\cite{HV-pythia} implementation in \PYTHIA[8.230], which handles the showering and hadronization of the dark quarks within the DS. Mediator masses ranging from $1~\TeV$ to $2.2~\TeV$ are used, with the corresponding pair-production cross-sections estimated using calculations from Supersymmetry, where stop--antistop pairs are used as a proxy for the dark mediators. The cross-sections are calculated at approximate next-to-next leading-order and next-to-next leading-logarithm (NNLOapprox + NNLL) using \textsc{NNLL-Fast}~\cite{NNLL-Fast,NNLLresum}. Cross-section values are multiplied by a factor of $N_{c}$, for the $N_{c} = 3$ model this gives values in the range from $20.5~\mathrm{fb}$ at $M_{\phi} = 1~\TeV$ to $0.015~\mathrm{fb}$ at $M_{\phi} = 2.2~\TeV$. Theoretical uncertainties corresponding to parton distribution functions, QCD scale, and resummation of Coulomb corrections are also produced using \textsc{NNLL-Fast} for each mediator mass.

Benchmark signal models are produced with $N_{c} = 3, N_{f} = 7$, following Ref.~\cite{SSW} to ensure a QCD-like DS. For a given $N_{c}$, there are a range of values for $N_{f}$, such that the running of the dark strong coupling ($\alpha_{d}$) is still asymptotically free, $N_{f} = 7$ is the centre of that range.

The dark confinement scale is used to set the masses of the DS particles with the following scheme: $\Lambda_{d} = m_{Q_{d}} = 2~m_{\pi_{d}} = \frac{m_{\rho_{d}}}{2}$. This forces the heavier dark mesons such as the dark rho to decay into dark pions ($\rho_{d} \to \pi_{d}\pi_{d}$), which occurs within the DS and therefore proceeds quickly. The subsequent dark pion decay to SM down quarks ($\pi_{d} \to d\bar{d}$) proceeds via $t$-channel exchange of a dark mediator. The decay is slow due to the large mass difference between the $\phi$ and $\pi_{d}$, giving rise to the macroscopic lifetimes of the dark pions. The production of dark baryons, which are suppressed by a factor of $1/N^{2}_{c}$~\cite{Cohen_2015}, or dark mesons other than the $\rho_{d}$ and $\pi_{d}$, is not considered.

Three choices of $\Lambda_{d}$ are used, corresponding to $m_{\pi_{d}}$ values of $5, 10,~\mathrm{and}~20~\GeV$. These values ensure that the decay of the dark pion into jets is valid, since lower mass $\pi_{d}$ would decay to distinct hadronic final states~\cite{Gross_Wilczek,Politzer}. For dark pion masses much larger than $\mathcal{O}(10)~\GeV$, the reduced phase space for dark pion production leads to a more diffuse spray of few dark pions giving a displaced jet signature, while this analysis targets the emerging jets signature with a collimated spray of many dark pions.

The dark pion lifetime ($c\tau_{\pi_{d}}$), as a function of the other phenomenological parameters of the model, is given by~\cite{SSW}:
\begin{equation}
c\tau = \frac{80~\mathrm{mm}}{\kappa^{4}} \Big(\frac{2~\GeV}{f_{\pi_d}}\Big)^{2} \Big(\frac{100~\MeV}{m_{q}^{\mathrm{SM}}}\Big)^{2} \Big(\frac{2~
\GeV}{m_{\pi_{d}}}\Big)\Big(\frac{M_{\phi}}{1~\TeV}\Big)^{4}
\end{equation}
where $\kappa$ is the Yukawa coupling between the dark mediator and the quarks, $f_{\pi_d}$ is the dark pion decay constant (analogous to the pion decay constant in the SM), and $m_{q}^{\mathrm{SM}}$ is the mass of the SM quark which the mediator couples to (in this case the SM down quark). A consistent set of dark pion lifetimes can be set for each choice of $M_{\phi}$ or $m_{\pi_{d}}$ by varying the Yukawa coupling. Samples are produced for proper decay lengths between $1~\mathrm{mm}$ and $1000~\mathrm{mm}$, such that the majority of the dark pions decay within the tracking detectors. Table~\ref{tab:params} gives a summary of the model parameters used in the signal MC samples for this analysis.

\begin{table}[h]
\caption{Signal model parameters used to give the model grid for the analysis.}
\label{tab:params}
\centering
\begin{tabular}{ c | c }
\hline \hline
$M_{\phi}$ [$\TeV$] & 1.0, 1.4, 1.6, 1.8, 2.0, 2.2 \\
\hline
$m_{\pi_{d}}$ [$\GeV$] & 5, 10, 20\\
\hline
$c\tau_{\pi_{d}}$ [mm] & 1,  2,  5,  20,  75,  150, 300,  500, 1000 \\
\hline
\hline
\end{tabular}
\end{table}

\subsection{Background Simulation}

While the background estimation strategy described in Section~\ref{sec:background} is fully data-driven, background MC samples are produced to optimise the analysis methodology.

The multijet QCD background~\cite{ATL-PHYS-PUB-2019-017} was generated using \PYTHIA[8.230] with dijet production matrix elements calculated at LO. The \NNPDF[2.3lo] PDF set is used for matrix element generation, showering, and multi-parton interactions, with the A14 set of parameters.

Samples of $t\bar{t}$ events were used to determine their contribution to the background. These were produced at next-to-leading-order (NLO) with \textsc{Powheg Box v2}~\cite{Nason:2004rx,Frixione:2007vw,Alioli:2010xd} using $h_{\mathrm{damp}} = 1.5~m_{t}$~\cite{	ATL-PHYS-PUB-2020-023}, then interfaced with \PYTHIA[8.230] for parton showering and hadronization. The decays of hadrons containing $b$ and $c$ quarks common in $t\bar{t}$ decays are performed by \textsc{EvtGen 1.6.0}~\cite{Lange:2001uf}. The contribution of $t\bar{t}$ events was less than $0.1\%$ in the search region after the analysis selections (described in Section~\ref{sec:selection}), and therefore are not considered in the background estimate.
\section{Event Reconstruction}
\label{sec:reconstruction}

The standard ATLAS track reconstruction in Run 2 uses a combinatorial Kalman filter~\cite{KalmanFilter} to create tracks from energy deposits in the ID. It is optimised for reconstructing tracks from prompt charged particles with $\pt > 500~\MeV$ and transverse impact parameters ($d_{0}$) less than $10~\mathrm{mm}$, but is largely insensitive to tracks which are displaced from the interaction point (IP) as a result of the decay of neutral long-lived particles. In order to increase the tracking acceptance to displaced signatures, a secondary reconstruction for large impact parameter tracks is performed on hits in the ID not associated with tracks from the standard track reconstruction. This dedicated Large Radius Tracking (LRT)~\cite{ATL-PHYS-PUB-2017-014} greatly relaxes the impact parameter selection to $|d_{0}| < 300~\mathrm{mm}$, with a stricter momentum requirement of $\pt > 900~\MeV$.

Events are required to have a primary vertex, associated to the hard scatter interaction, which must have at least two tracks associated with it~\cite{ATL-PHYS-PUB-2015-026}. Because there may be several interaction vertices within the event, the primary interaction vertex is selected as the one with the largest $\sum \pt^{2}$ of all associated standard tracks.

Secondary displaced vertices (DVs) are reconstructed using a dedicated algorithm seeded with tracks from both the standard and LRT track collections~\cite{ATL-PHYS-PUB-2019-013}. In order to mitigate the contribution of background DVs from SM processes, material interactions, and fakes, a series of analysis-specific requirements were defined which apply successive selections to the DVs. These aim to reject background DVs while maintaining high efficiency of selecting signal DVs. The first selections are on the radial and longitudinal position of the DV, restricting DVs to where the LRT tracking is efficient, $r_{\mathrm{DV}} < 300~\mathrm{mm}$ and $|z_{\mathrm{DV}}| < 300~\mathrm{mm}$. Next, to reduce the contribution from long-lived neutral kaons ($K_{S}^{0}$), the DVs are required to have $m_{\mathrm{DV}} > 0.7~\GeV$, where the mass is computed as the invariant mass of the sum of the four-momenta of all tracks associated to the DV, where each track is assumed to be from a SM charged pion. The effect of random track crossings being reconstructed as DVs is limited by applying requirements on the minimum of the impact parameters of all tracks matched to the DV ($\min|d_{0}| < 10~\mathrm{mm}$, $\min|z_{0}| < 100~\mathrm{mm}$), and the transverse momenta ($\pt^{\mathrm{DV}} > 2.5~\GeV$) of the reconstructed DVs. Selections on the associated uncertainties of the impact parameters ($\sigma_{d_{0}} < 0.25~\mathrm{mm}$, $\sigma_{z_{0}} < 2.25~\mathrm{mm}$) help to remove DVs reconstructed from LRT fakes. Finally, a map of the ID material obtained with Run 2 data~\cite{PERF-2015-07} is used to veto DVs which are located in regions of known detector material.

Jets are reconstructed using energy deposits in the calorimeters. Topological clusters~\cite{PERF-2014-07, PERF-2016-04} are reconstructed into jets using the anti-$k_{t}$ algorithm using a distance parameter of $R = 0.4$, and are calibrated at the electromagnetic scale. In order to be counted in the jet collection, jets are required to have $\pt > 50~\GeV$ and $|\eta| < 2.7$.

Additionally, jets containing $b$-hadrons ($b$-jets) are tagged using the DL1r algorithm~\cite{FTAG-2019-07}, passing the $85\%$ efficiency working point, and are used to estimate the contribution of heavy-flavour within the multijet QCD background. In order to keep jets with associated displaced tracks, which are expected with the emerging jets topology, no attempt is made to match jets to the primary vertex.

\section{Event Selection}
\label{sec:selection}

Events are selected using an unprescaled trigger which requires at least 4 jets with a $\pt$ threshold of $120~\GeV$~\cite{TRIG-2019-04}. The trigger has a Level 1 requirement of three jets that each deposit at least 50 GeV of transverse energy, within a region of $0.8 \times 0.8$ in $\eta - \phi$ space~\cite{TRIG-2012-01,TDAQ-2019-01} in the calorimeters. As described above, the analysis uses a set of reconstruction techniques which were non-standard and computationally expensive in Run 2. Thus, this special reconstruction was only run on events which passed the baseline topology requirements of at least four jets, each with transverse momentum $\pt > 120~\GeV$ and $|\eta| < 2.4$. Events matching this topology are selected using jets before calibration, which is applied after reconstruction.

A dedicated emerging jets trigger was implemented in Run 3, which allowed for the analysis in Ref.~\cite{EXOT-2022-31} to increase their sensitivities to lower jet multiplicity topologies. However, in Run 2, only the 4-jet trigger provided sufficient background rejection to allow for sensitivity to the emerging jets models.

Requirements are imposed on the four highest $\pt$ jets within the event to ensure an emerging jet-like topology. The scalar sum $\pt$ of these jets ($H_{\mathrm{T}}$) is required to be greater than $1~\TeV$. At least one of the four highest $\pt$ jets is also required to have a well-defined prompt track fraction (PTF):
\begin{equation}
\mathrm{PTF} = \frac{1}{\pt^{\mathrm{Jet}}} \sum_{i\in \mathrm{tracks}} \Big(p_{\mathrm{T}, i} \Big| \frac{|d_{0, i}|}{\sigma_{d_{0}, i}} < 3\Big).
\end{equation}
For jets within $|\eta| < 2$, the PTF is computed if at least one standard track with $\pt^{\mathrm{Track}} \geq 2~\GeV$ can be matched within $\Delta R < 0.4$, even if it does not meet the promptness criterion of the $d_{0}$ significance.

Finally, events are required to have at least one displaced vertex, passing all the selections defined above, which is matched to a reconstructed jet, to ensure the event meets the key criterion of emerging jets. This jet-matching for DVs is done using a loosened angular distance criterion of $\Delta R < 0.6$ between the DV position and the jet axis. This looser selection is used to account for the potential of jet-splitting within the EJs, where certain combinations of model parameters result in more diffuse sprays of high energy dark pions, which can potentially be reconstructed as individual jets, rather than a single emerging jet. This effect both inflates the jet multiplicity ($N_{\mathrm{Jet}}$) of the event and decreases the displaced vertex multiplicity of the reconstructed jets.

\section{Machine Learning Classifier}
\label{sec:analysis}

The analysis uses a two-stage ML approach using boosted decision trees (BDTs) implemented with the Toolkit for Multivariate Analysis (\texttt{TMVA})~\cite{TMVA}. The first step trains a BDT on jet-level information to select EJ candidates, while the second step uses a BDT trained on event-level information, including the BDT scores of the four highest $\pt$ jets, to separate EJ signal events from SM multijet background. A cut-based version of the analysis, which is not optimized to the same degree as the ML strategy, is described in Appendix~\ref{sec:appendix} and included in HEPData to aid with reinterpretation of the results.

In order to avoid biasing the results, both signal and background events which pass the selections defined in Section~\ref{sec:selection} are separated into two equal sets. The first set is used only for training both stages of the BDT and the second is used for testing. While the two sets are statistically independent, their BDT response distributions are consistent within the MC statistical uncertainties, showing the classifier is not overtrained. Both the jet- and event-level BDTs are trained on all of the signal MC samples mixed together, with background events sampled equally from across the full $\pt$ spectrum of the multijet MC. The full training set comprises over 9 million events.

\begin{table}[h]
\caption{List of variables used in training for the jet-level boosted decision tree.}
\label{tab:jetVariables}
\centering
\begin{tabular}{ c | c }
\hline \hline
Observable & Description \\
\hline
$E_{\mathrm{Jet}}$ & Jet energy, measured directly from calorimeter deposits\\
$m_{\mathrm{Jet}} = \sqrt{\big(\sum_{i}E_{i}\big)^{2} - \big(\sum_{i}\vec{p}_{i}\big)^{2}}$ & Jet mass, invariant mass from the sum of topological clusters\\
$W_{\mathrm{Jet}} = \frac{1}{\pt^{\mathrm{Jet}}}\sum_{i}p_{\mathrm{T}, i}\Delta R_{i}$ & Jet width, $\pt$-weighted angular spread of topological clusters\\
$\eta_{\mathrm{Jet}}$ & Jet pseudorapidity \\
\hline
\hline
\end{tabular}
\end{table}

The jet-level BDT is trained on each of the four highest $\pt$ jets in the event, using the jet-level observables given in Table~\ref{tab:jetVariables}. After training, the individual jet-level BDT responses ($R_{\mathrm{Jet}}$) of the four highest $\pt$ jets are used as inputs to the event-level BDT.

\begin{table}[h]
\caption{List of variables used in training for the event-level boosted decision tree.}
\label{tab:eventVariables}
\centering
\begin{tabular}{ c | c }
\hline \hline
Observable & Description \\
\hline
$R_{\mathrm{Jet}}^{1-4}$ & BDT response of the four highest $\pt$ jets, individually\\
$H_{\mathrm{T}} = \sum_{j = 1}^{4} p_{\mathrm{T},j}$ & Scalar sum of the four highest $\pt$ jet transverse momenta\\
minPTF & The minimum PTF value in the four highest $\pt$ jets\\
$N_{\mathrm{Jet}}$ & Reconstructed jet multiplicity\\
$S_{\mathrm{T}}$ & Transverse Sphericity\\
\hline
\hline
\end{tabular}
\end{table}

In addition to the four highest $\pt$ jet BDT responses, the event-level BDT takes several observables which are used to help separate signal and background. The full list is summarized in Table~\ref{tab:eventVariables}. One of the observables, transverse sphericity ($S_{\mathrm{T}}$), is an isotropy measure~\cite{STDM-2019-02}, which characterizes the shape of the event. Background events, for which the majority of the event activity is concentrated in the two highest $\pt$ jets (often balanced in the transverse plane), are expected to have values of $S_{\mathrm{T}}$ close to 0. On the other hand, signal events tend to have a larger multiplicity of jets arranged more isotropically, leading to transverse sphericity values closer to 1.

Figure~\ref{fig:BDTresponse} shows the event-level BDT response observable ($R$) evaluated on background MC and a selection of signal models. In general, models which present as less SM-like are better separated from multijet background. For this reason, higher $M_{\phi}$ models tend to higher BDT responses, but the sensitivity becomes limited by the reduced cross-sections. With respect to the dark pion lifetime, only models with very small $c\tau_{\pi_{d}}$ are difficult to separate from the background, however sensitivity is also lost for very long lifetimes since more of the $\pi_{d}$ decay outside of the tracker acceptance to reconstruct a displaced vertex, so the higher BDT responses are seen in intermediate lifetime models. The BDT is largely insensitive to $m_{\pi_{d}}$, with the $10$ and $20~\GeV$ models being only slightly better separated than the $5~\GeV$ models.

An additional selection of $R > -0.2$ is applied to improve the background estimation discussed in Section~\ref{sec:background}. This requirement removes mainly background events, with at most $0.1\%$ impact on the signal efficiency.

\begin{figure}[h]
\centering
\includegraphics[width=0.7\textwidth]{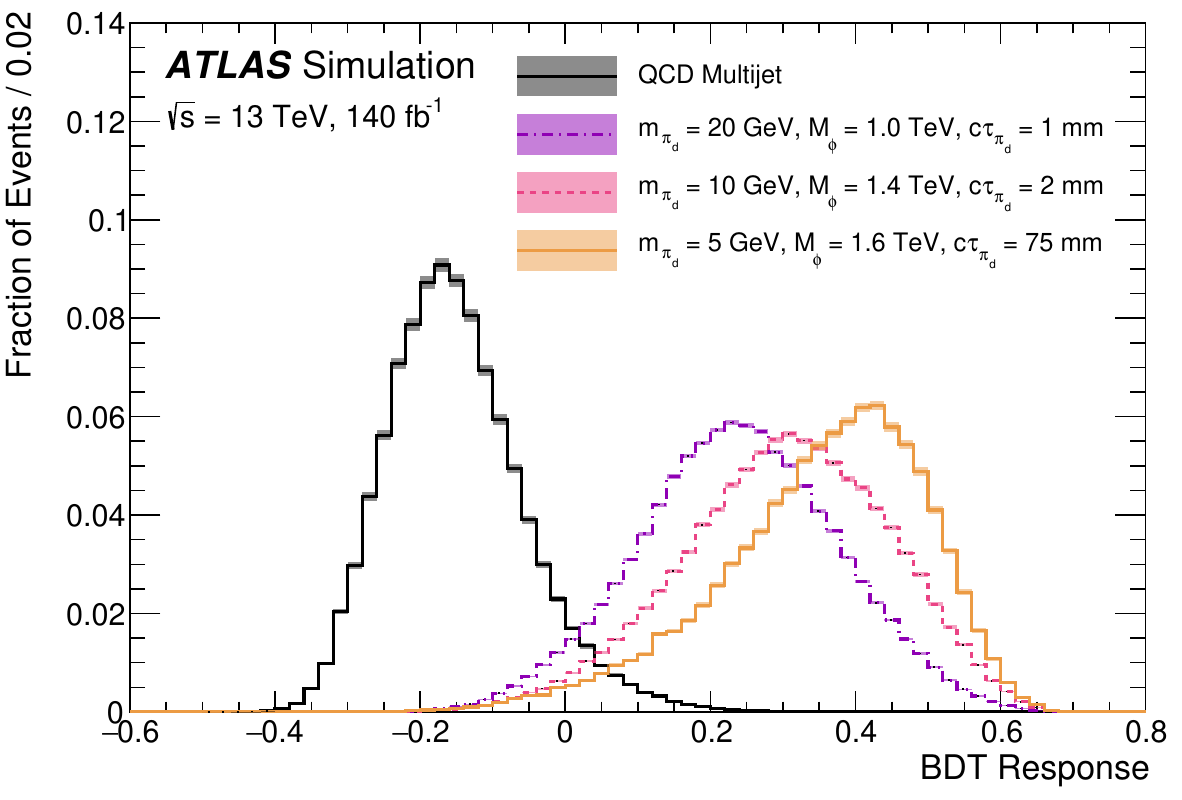}
\caption{Normalized event-level BDT response distribution for the QCD multijet background, with three selected signal samples for comparison.}
\label{fig:BDTresponse}
\end{figure}

\section{Background Estimation}
\label{sec:background}

The dominant background for this analysis is QCD multijet production. However, these processes are difficult to properly model in simulation, thus the background estimate is produced using a data driven technique modified from the ABCD method.

The ABCD method employs a 2D parameter space divided into four regions (labelled A, B, C, and D) by applying selection criteria on each of the two variables. If the background distributions of these variables are independent, the ratios of expected background yields are related by: $\frac{N_{A}}{N_{B}} = \frac{N_{C}}{N_{D}}$. If the plane is designed such that the background primarily populates regions B, C, and D (labelled control regions, CRs), while the signal is concentrated in region A (labelled signal region, SR), the background yield in the SR can be estimated as:

\begin{equation}
N_{A} = N_{B} \times \frac{N_{C}}{N_{D}}
\label{eq:ABCD}
\end{equation}

When performing the final statistical analysis, any signal contamination in the CRs is accounted for using a simultaneous fit, as described in Section~\ref{sec:result}.

The ABCD plane is defined using the event-level BDT response and the number of jet-matched displaced vertices in the event ($N_{\mathrm{DV}}$). The distributions of events in the ABCD plane are shown for data and an example signal model in Figure~\ref{fig:ABCDplanes}. The SR is defined by $R \geq 0.2$ and $N_{\mathrm{DV}} \geq 2$. Control regions B and C are defined by inverting one of the two selections, and region D is defined by inverting both selections. For signal models with $M_{\phi} = 1.6~\TeV$ and $m_{\pi_{d}} = 10~\GeV$, the full set of SR selections have efficiencies which vary between $1.4\%$ for $c\tau_{\pi_{d}} = 1000~\mathrm{mm}$ and $72\%$ for $c\tau_{\pi_{d}} = 20~\mathrm{mm}$.

\begin{figure}[h]
\centering
\includegraphics[width=0.49\textwidth]{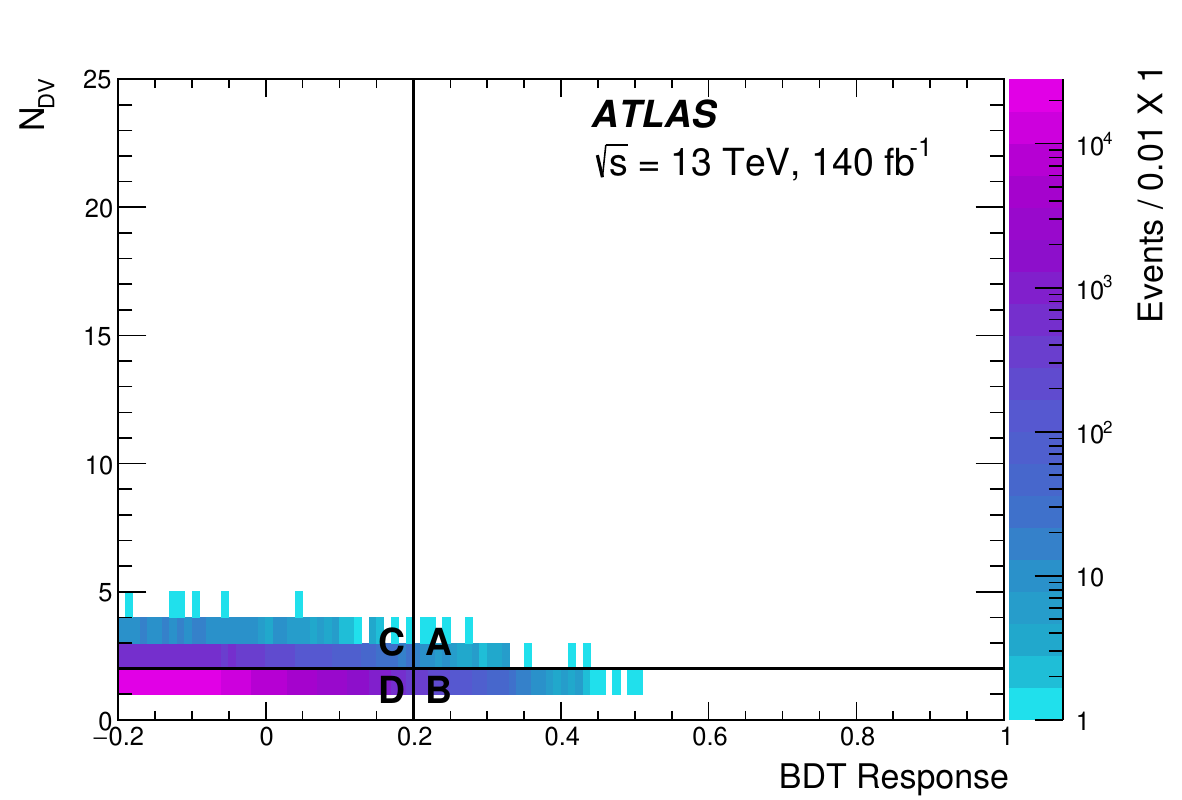}
\includegraphics[width=0.49\textwidth]{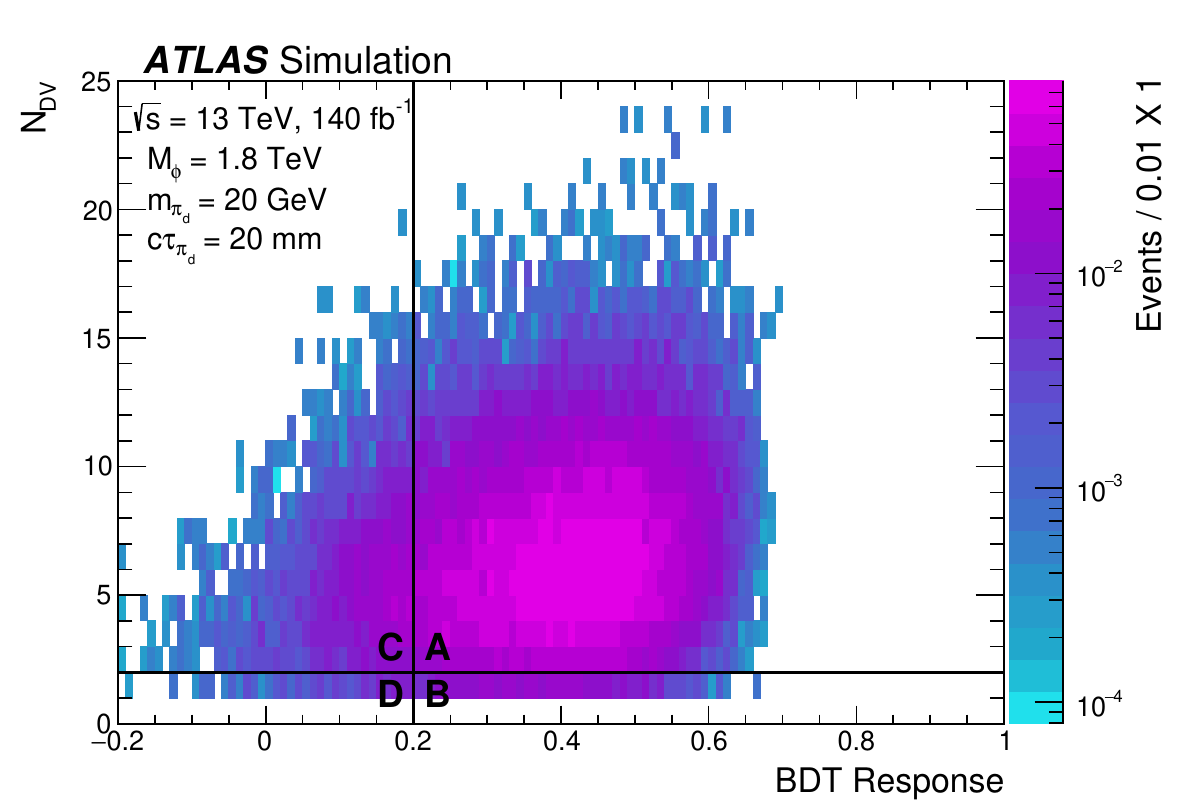}
\caption{Distributions of events in the ABCD plane for (left) Run 2 data and (right) an example signal model with parameters: $m_{\pi_{d}} = 20~\GeV$, $M_{\phi} = 1.8~\TeV$, and $c\tau_{\pi_{d}} = 20~\mathrm{mm}$. The signal events are scaled to the pair-production cross-section for $pp \to \phi^{\dagger}\phi$ at $M_{\phi} = 1.8~\TeV$, and normalized to the integrated luminosity of the dataset.}
\label{fig:ABCDplanes}
\end{figure}

To account for non-negligible correlations between the two variables, CRs C and D are subdivided and the ratio $\frac{N_C}{N_D}$ is calculated in each subdivision. This is then fit with a two-parameter linear function which can be extrapolated into the SR. This function then replaces the ratio in Equation~\ref{eq:ABCD}, to become:

\begin{equation}
N_{A} = (p_{0} \cdot R_{B} + p_{1}) \times N_{B}
\end{equation}
where $p_{0}$ and $p_{1}$ are the slope and y-intercept of the linear function, respectively, and $R_{B}$ is the event-weighted mean BDT response value within region B. The statistical uncertainty on the background estimation is a combination of the propagated statistical uncertainties from the fit parameters and the statistical uncertainty from the counts in region B. This corresponds to an uncertainty of $5\%$ on the final background estimate.

\begin{figure}[h]
\centering
\includegraphics[width=0.7\textwidth]{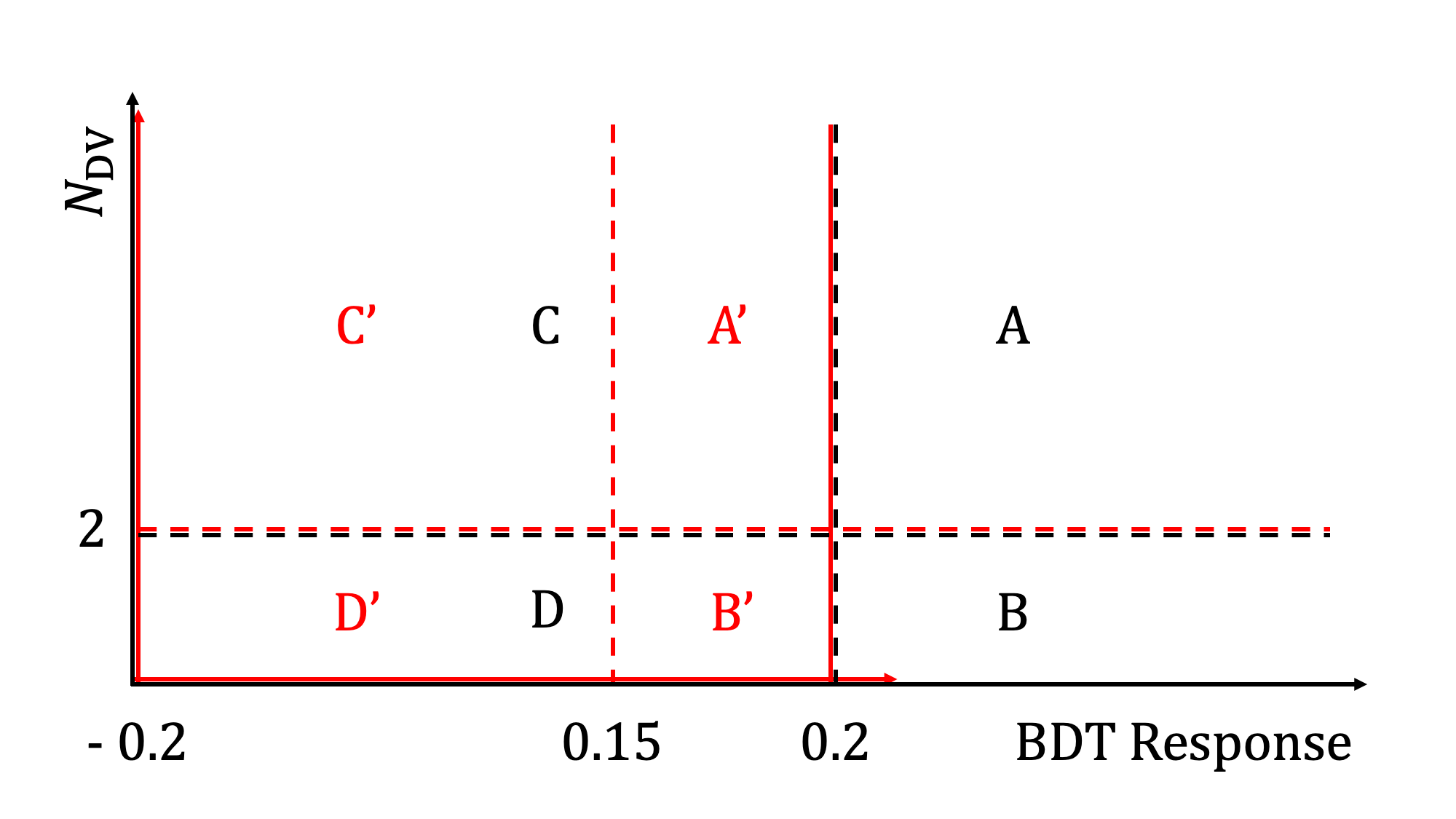}
\caption{Schematic diagram of the Shifted-ABCD VR constructed in CRs C and D. The four new regions are shown in red, and labeled A', B', C', and D'. The actual ABCD plane regions are shown in black for comparison.}
\label{fig:ABCDprime}
\end{figure}

In order to validate the background estimation method, two validation regions (VRs) are constructed. The Low-$H_{\mathrm{T}}$ VR is defined by inverting the $H_{\mathrm{T}}$ requirement used in the event selection to be $H_{\mathrm{T}} < 1~\TeV$, while keeping the other selections in place. This allowed for the construction of another ABCD plane using a background enriched region where data could be analysed without biasing the method. The Low-$H_{\mathrm{T}}$ VR is further restricted to be between $775~\GeV \leq H_{\mathrm{T}} < 1000~\GeV$ to ensure good modelling, which rapidly deteriorates at lower $H_{\mathrm{T}}$ values. Since the $H_{\mathrm{T}}$ requirement is inverted, signal contamination in the Low-$H_{\mathrm{T}}$ VR is below $0.001\%$ for all signal models.

The second VR is constructed using CRs C and D from the ABCD plane, which have the largest contribution of background events, and the smallest signal contamination. The Shifted-ABCD VR, shown in Figure~\ref{fig:ABCDprime}, is defined with an additional selection on the BDT response, along with the same selection on $N_{\mathrm{DV}}$ used in the actual ABCD plane, to define four new regions. The pseudo-signal region A' is defined for $0.15 \leq R < 0.2$ and $N_{\mathrm{DV}} \geq 2$, with the corresponding primed CRs defined by inverting one or both of the selections. Since it is much closer to the actual SR, signal contamination is slightly higher in region A' than in the Low-$H_{\mathrm{T}}$ VR, however, all models still contribute less than $10\%$ of the total MC event yield.

The predicted and observed counts in the SRs of both the Low-$H_{\mathrm{T}}$ VR and the Shifted-ABCD VR are given in Table~\ref{tab:ABCDclosure}. The modified ABCD method is shown to predict background contributions which agree with the observed yields within their respective uncertainties. The per-bin closure in both validation regions was also measured to be within $1\sigma$ for all bins.

\begin{table}[h]
\caption{Predicted and observed counts in the A' regions of the Low-$H_{\mathrm{T}}$ VR and the Shifted-ABCD VR, using the modified ABCD method. Quoted uncertainties are statistical only.}
\label{tab:ABCDclosure}
\centering
\begin{tabular}{ c | c c }
\hline \hline
Region & Low-$H_{\mathrm{T}}$ VR & Shifted-ABCD VR \\
\hline
Predicted Counts & $25 \pm 3$ & $215 \pm 8$ \\
Observed Counts & $24$ & $209$ \\
\hline
\hline
\end{tabular}
\end{table}

The relative difference of $3\%$ between the predicted and observed background yields in region A' of the Shifted-ABCD VR is used as a systematic uncertainty for potential non-closure of the modifed ABCD method in the actual SR. Both the statistical uncertainty and the systematic uncertainty on the background estimation method are included separately in the statistical analysis (discussed in Section~\ref{sec:result}).

The closure of the modified ABCD method was also tested using QCD multijet simulation. Good agreement of the predicted and observed event yields in the SR was found, with per-bin closure within $2\sigma$ for all bins in the signal region. This closure held both inclusively, and when tested on heavy-flavour QCD production by selecting events where at least one of the four highest $\pt$ jets is $b$-tagged using either the DL1r classifier or truth-level information.

\section{Signal Systematic Uncertainties}
\label{sec:systematics}

Experimental uncertainties relating to instrumentation or reconstruction are assigned to the modelling of simulated signal samples. These systematic uncertainties include contributions from jets, tracks and displaced vertices, luminosity, as well as a dedicated uncertainty for the prompt track fraction.

The systematic associated with jets has contributions from the jet energy scale (JES), jet energy resolution (JER)~\cite{JETM-2018-05}, and jet mass scale (JMS)~\cite{JETM-2018-02}. Jets are varied within the uncertainties for each of these contributions and the effects are propagated to the final signal yield. The resulting uncertainty depends primarily on the mediator mass of the signal sample, with dark pion lifetime having a larger effect at smaller mediator masses. The combined systematic varies between $1\%$ for $M_{\phi} = 2.2~\TeV$ and $13\%$ for $M_{\phi} = 1~\TeV$.

Uncertainties from tracks and displaced vertices are derived from a combination of sources to account for both the standard and large-radius tracks. The uncertainty from LRT is derived from comparisons of events consistent with $K_{S}^{0} \to \pi^{+} \pi^{-}$ between simulation and data, and follows a prescription defined in Ref.~\cite{EXOT-2018-57}. To assess the impact, the track collection is varied according to these differences, summed in quadrature with the $1.7\%$ systematic uncertainty associated with standard tracking~\cite{ATL-PHYS-PUB-2015-051}, and then vertex reconstruction is repeated using the altered set of tracks. These changes are eventually propagated to track and DV observables, in particular $N_{\mathrm{DV}}$, which gives the dominant source of uncertainty. In this case, the dark pion lifetime has the biggest impact on the size of the uncertainty, which varies between $0.5\%$ for $c\tau_{\pi_{d}} = 2~\mathrm{mm}$ and $39\%$ for $c\tau_{\pi_{d}} = 1000~\mathrm{mm}$. The mediator mass also has an impact on the systematic uncertainty, since lower $M_{\phi}$ result in lower multiplicities of dark pions.

The dedicated systematic uncertainty to cover the modelling of PTF is determined by comparing the distribution of minPTF in data and background MC using the Low-$H_{\mathrm{T}}$ VR region. Per-event weights are created based on the differences, which are then used to reweight the signal samples. The final uncertainty is taken as the difference in the signal yield and varies between $1\%$ and $12\%$, where the combination of lower $M_{\phi}$ and $c\tau_{\pi_{d}}$ result in larger uncertainties.

Pile-up conditions during different data-taking periods are modelled by reweighting simulated events, such that the average number of interactions per bunch crossing matches the observed distribution in data. These contributions depend on the inelastic $pp$ cross-section, so the difference between predicted and measured values~\cite{STDM-2015-05} are propagated as a systematic variation to reweighted events. The pile-up uncertainty varies between $0.3\%$ for $M_{\phi} = 1~\TeV$ and $5\%$ for $M_{\phi} = 2~\TeV$.

Signal MC events are scaled to the integrated luminosity of the dataset, and are therefore subject to the uncertainty associated with the measurement of the total luminosity. For the full Run 2 dataset, the uncertainty is $0.8\%$~\cite{DAPR-2021-01}, obtained using the LUCID-2 detector~\cite{LUCID2}.

Theoretical uncertainties relating to the production of the dark mediators are applied to the theoretical cross-sections used for limit-setting in Section~\ref{sec:result}. These uncertainties are produced using \textsc{NNLL-Fast}~\cite{NNLL-Fast,NNLLresum} along with the corresponding cross-sections, and are roughly proportional to the mediator mass. Values range from $11\%$ at $M_{\phi} = 1~\TeV$ to $27\%$ for $M_{\phi} = 2~\TeV$.

The dominant source of uncertainty varies between the different signal models, but in most cases the tracking and vertexing systematic is the leading contribution.
\section{Results}
\label{sec:result}

The predicted background yield in the SR, determined from the extrapolated fit of the modified ABCD method (described in Section~\ref{sec:background}), and corresponding observed yield is given in Table~\ref{tab:FinalYield}. No excess above the data-driven background estimate is observed.

\begin{table}[h]
\caption{Predicted number of background events, including statistical and systematic uncertainties, and observed event yield in the SR.}
\label{tab:FinalYield}
\centering
\begin{tabular}{ c | c  }
\hline \hline
\multicolumn{2}{c}{Event Yield in the SR} \\
\hline
Prediction & $112 \pm 6~\mathrm{(stat.)} \pm 4~\mathrm{(syst.)}$ \\
Observed Yield & $103$ \\
\hline
\hline
\end{tabular}
\end{table}

In the absence of excess data events, a hypothesis test is performed using a simultaneous likelihood fit which incorporates the methodology of the modified ABCD method, and accounts for any signal leakage into the CRs. Signal and background systematic uncertainties are included in the fit as Gaussian nuisance parameters.

\begin{figure}[h]
\centering
\includegraphics[width=0.65\textwidth]{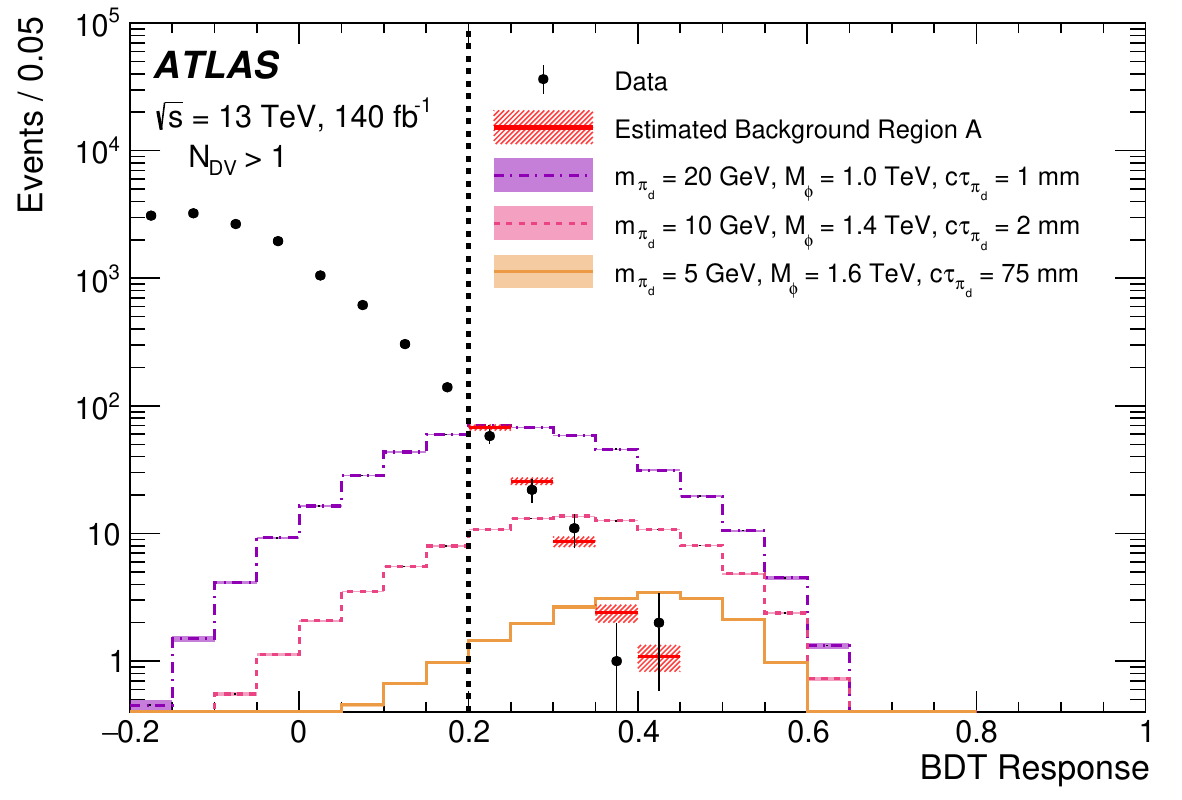}
\caption{Comparison of the data with $N_{\mathrm{DV}} > 1$ and the estimated background in the SR using the modified ABCD method. The division between the SR and CR is shown by the vertical dashed line. In the final fit, the bins with $R > 0.4$ are combined into a single bin. Three selected signal samples are included for comparison.}
\label{fig:unblindedBDT}
\end{figure}

The likelihood fit is constructed with the product of Poisson probabilities of observing the event yield in each region $i \in \{A,B,C,D\}$ of the ABCD plane, subdivided into $j$ bins across the BDT response ($N^{\mathrm{Obs}}_{i,j}$) in order to perform the linear fit across regions C and D, which is then extrapolated into the regions A and B. The CRs C and D are divided into 8 bins across the BDT response, while the SR and CR B are divided into 4-5 bins, where the number of bins is adjusted to ensure there are sufficient data events in each bin of region B. The expected number of events in a given bin of region A ($N^{\mathrm{Exp}}_{A,j}$) is then given by:
\begin{equation}
N^{\mathrm{Exp}}_{A,j} =  (p_{0} \cdot R_{j} + p_{1}) \times N_{B, j}^{\mathrm{Exp}} + \mu\cdot N_{A,j}^{\mathrm{Sig}}
\end{equation}
where $p_{0}$ and $p_{1}$ are again the slope and y-intercept, respectively, of the linear fit extracted from the subdivided regions C and D. $R_{j}$ is the event weighted mean value of the BDT response within the $j$ bin of region B. The signal strength parameter, $\mu$, is fit simultaneously across the entire ABCD plane, and $N_{A,j}^{\mathrm{Sig}}$ is the expected signal contribution in each bin of A, taken from signal MC. The agreement between the extrapolated background estimate, as determined by the likelihood fit, and the observed data in the SR is shown in Figure~\ref{fig:unblindedBDT}.  Only data and MC from the SR and region C are shown, since events are required to have $N_{\mathrm{DV}} > 1$.

The full likelihood function is given by:
\begin{equation}
\mathcal{L}\big(\mu, \theta\big) = \prod_{i \in \{A, B, C, D\}} \Big( \prod_{j \in i} \mathit{P} \big(N^{\mathrm{Obs}}_{i,j} | N^{\mathrm{Exp}}_{i,j} (p_{0}, p_{1}, \mu)\big) \Big) \Big( \mathit{G}\big( \alpha^{\mathrm{Sig}}_{i}, \sigma^{\mathrm{Sig}} | 0, 1\big) \Big) \times \prod_{k = 1}^{3} \mathit{G}\big( \alpha_{k}, \sigma_{k} | 0, 1\big)
\end{equation}
where $\mathit{P}$ represents the Poisson probabilities, and $\mathit{G}$ are the Gaussian constraints on the systematic uncertainties, with $\alpha$ and $\sigma$ the mean and standard deviations, respectively. The set of free parameters, other than the signal strength, is given by $\theta = \{p_{0}, p_{1}, N^{\mathrm{Exp}}_{B,j}, N^{\mathrm{Exp}}_{D,j}, \alpha^{\mathrm{Sig}}_{i}, \alpha_{k}\}$.

\begin{figure}[h]
\centering
\subfloat[]{
\includegraphics[width=0.48\textwidth]{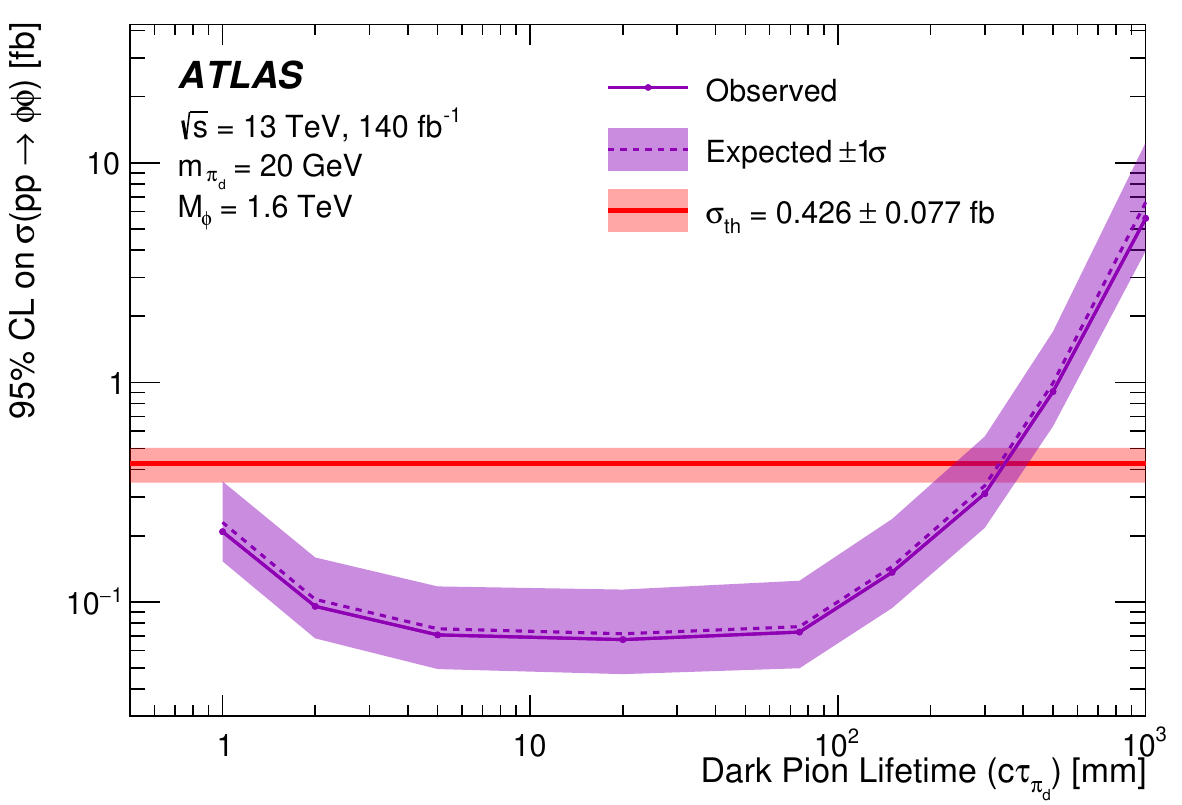}
\label{fig:mPi20_MX1600}
}
\subfloat[]{
\includegraphics[width=0.48\textwidth]{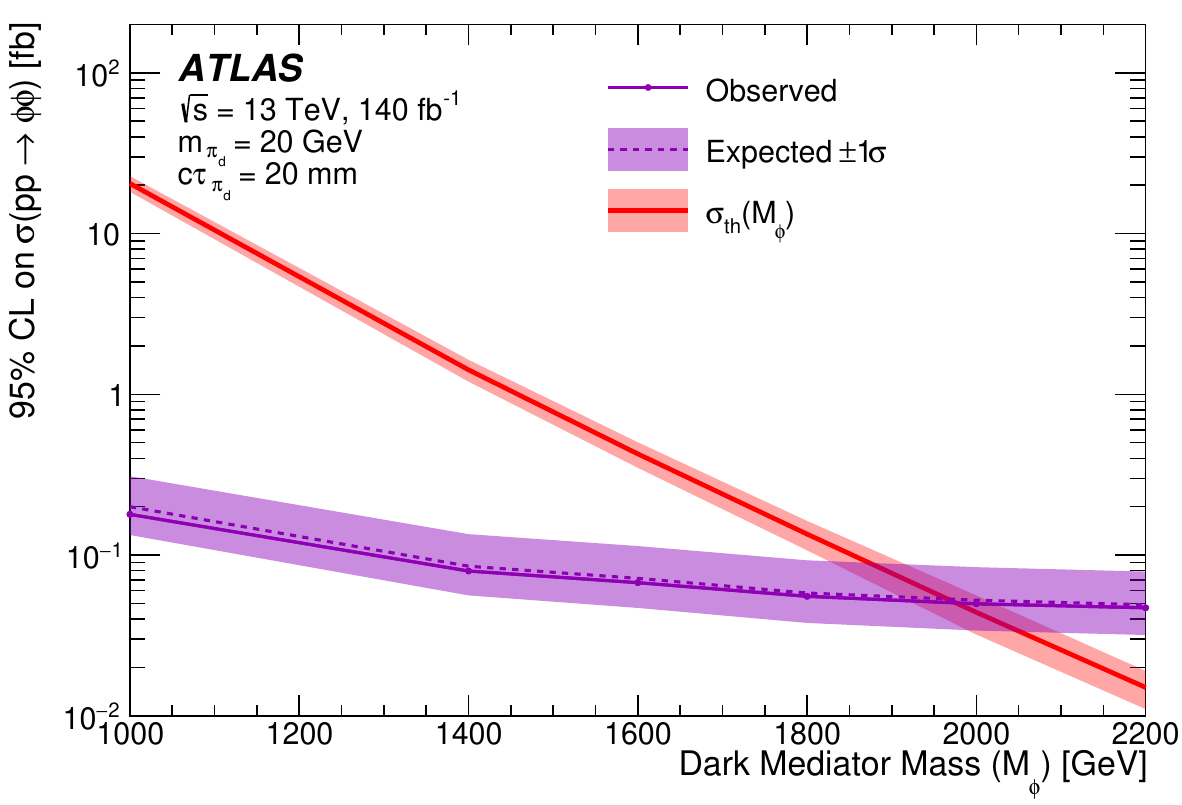}
\label{fig:mPi20_ct20}
}\\
\subfloat[]{
\includegraphics[width=0.48\textwidth]{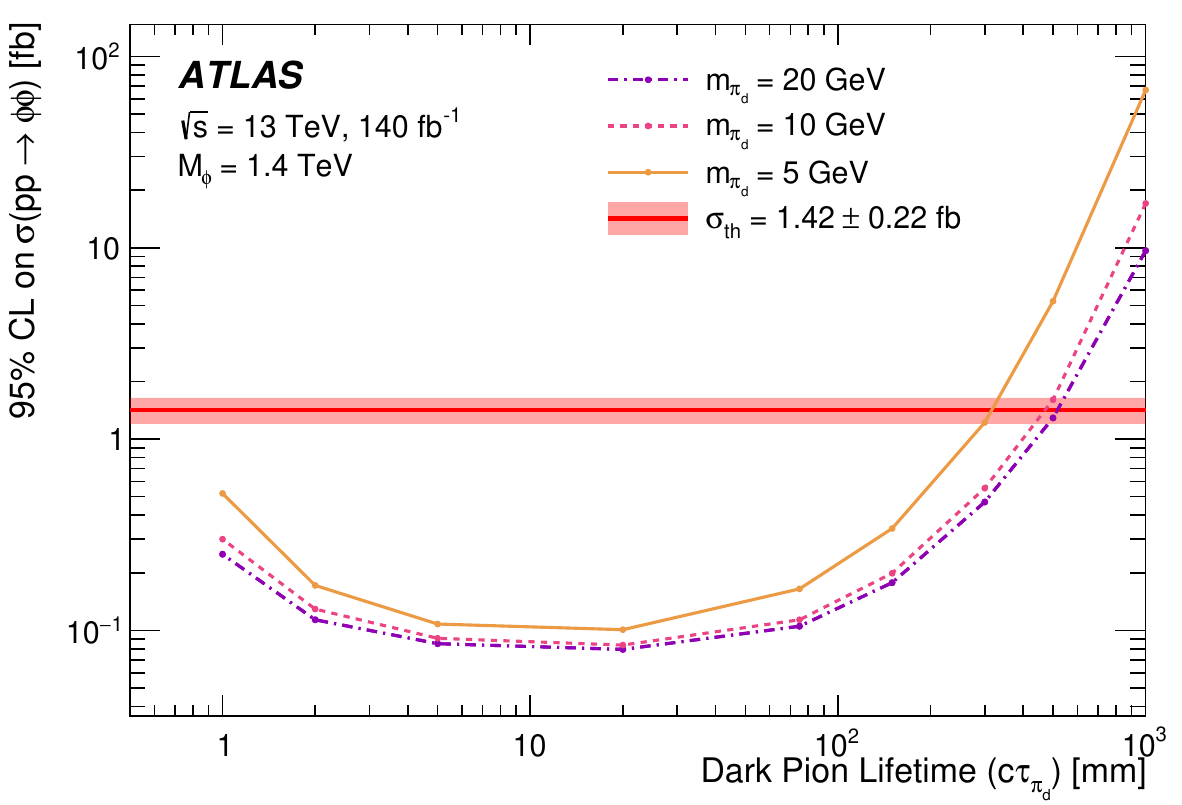}
\label{fig:MX1400}
}
\subfloat[]{
\includegraphics[width=0.48\textwidth]{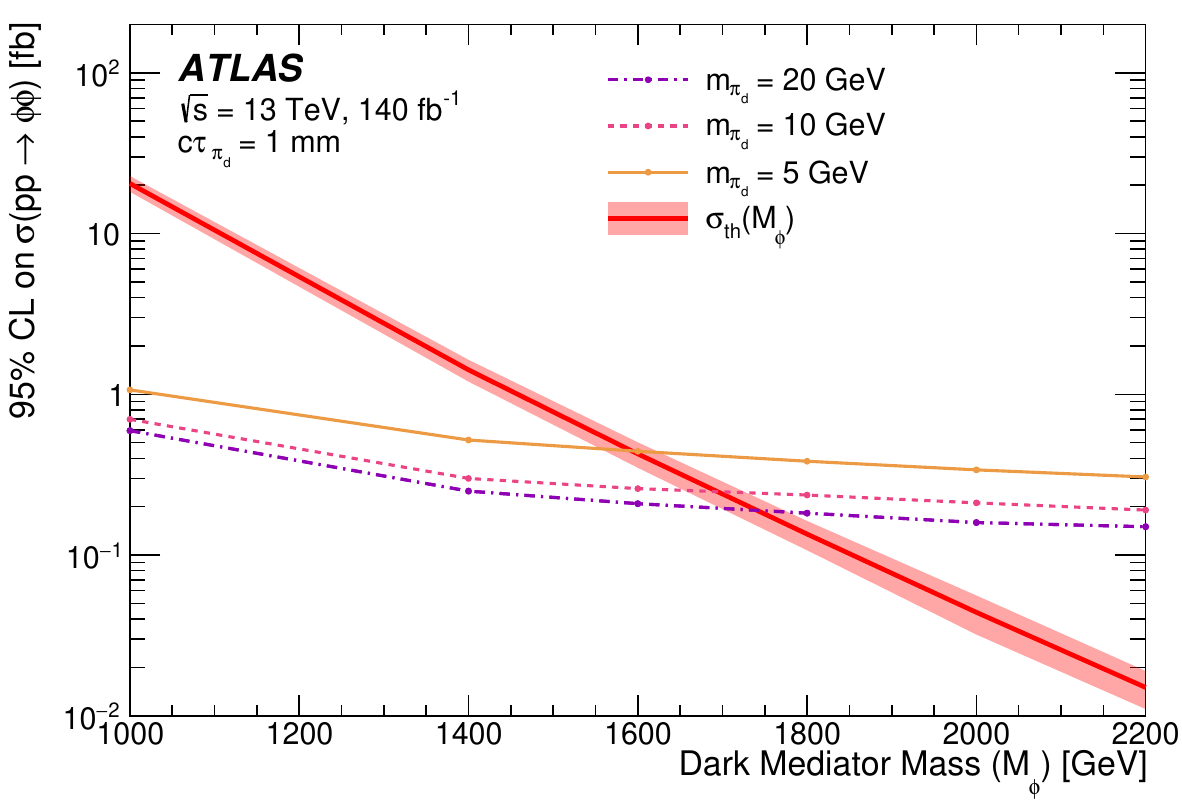}
\label{fig:ct1}
}
\caption{$95\%$ CL upper limits as a function of (left) $c\tau_{\pi_{d}}$ and (right) $M_{\phi}$. The upper plots show the expected and observed limits on $\sigma(pp \to \phi^{\dagger}\phi)$ for $m_{\pi_{d}} = 20~\GeV$: \protect\subref{fig:mPi20_MX1600} using $M_{\phi} = 1.6~\TeV$ and \protect\subref{fig:mPi20_ct20} using $c\tau_{\pi_{d}} = 20~\mathrm{mm}$. The lower plots show a comparison of the observed limits for all three dark pion masses: \protect\subref{fig:MX1400} using $M_{\phi} = 1.4~\TeV$, and \protect\subref{fig:ct1} using $c\tau_{\pi_{d}} = 1~\mathrm{mm}$. The mediator mass-dependent theoretical cross-section is given with the band corresponding to the uncertainty from \textsc{NNLL-Fast}.}
\label{fig:CLs_MX_ct}
\end{figure}

The four signal systematic uncertainties (jet, track and vertex, pile-up, and PTF) are combined, but kept decorrelated across the four ABCD regions; the standard deviation of the Gaussian constraint ($\sigma^{\mathrm{Sig}}$) is set to the same value across the four regions (set by the combined systematic uncertainty), but the mean ($\alpha^{\mathrm{Sig}}$) is allowed to float independently in each region. This allows for different systematic uncertainty values in each region of the ABCD plane. The three other systematic uncertainties, one for the luminosity uncertainty, and two for the background estimate uncertainty (which are applied only to region A), are included as individual nuisance parameters. In these cases, the Gaussian terms are correlated across the full ABCD plane giving only a single value for the mean of each term, while the standard deviations are set by the corresponding systematic uncertainties.

The test statistic is defined using the profile likelihood ratio whose probability distribution is determined with the asymptotic approximation~\cite{Cowan_2011}, and which gives a signal strength that is strictly positive. The likelihood implementation is handled by the \texttt{xRooFit} package.

\begin{figure}[h]
\centering
\subfloat[]{
\includegraphics[width=0.48\textwidth]{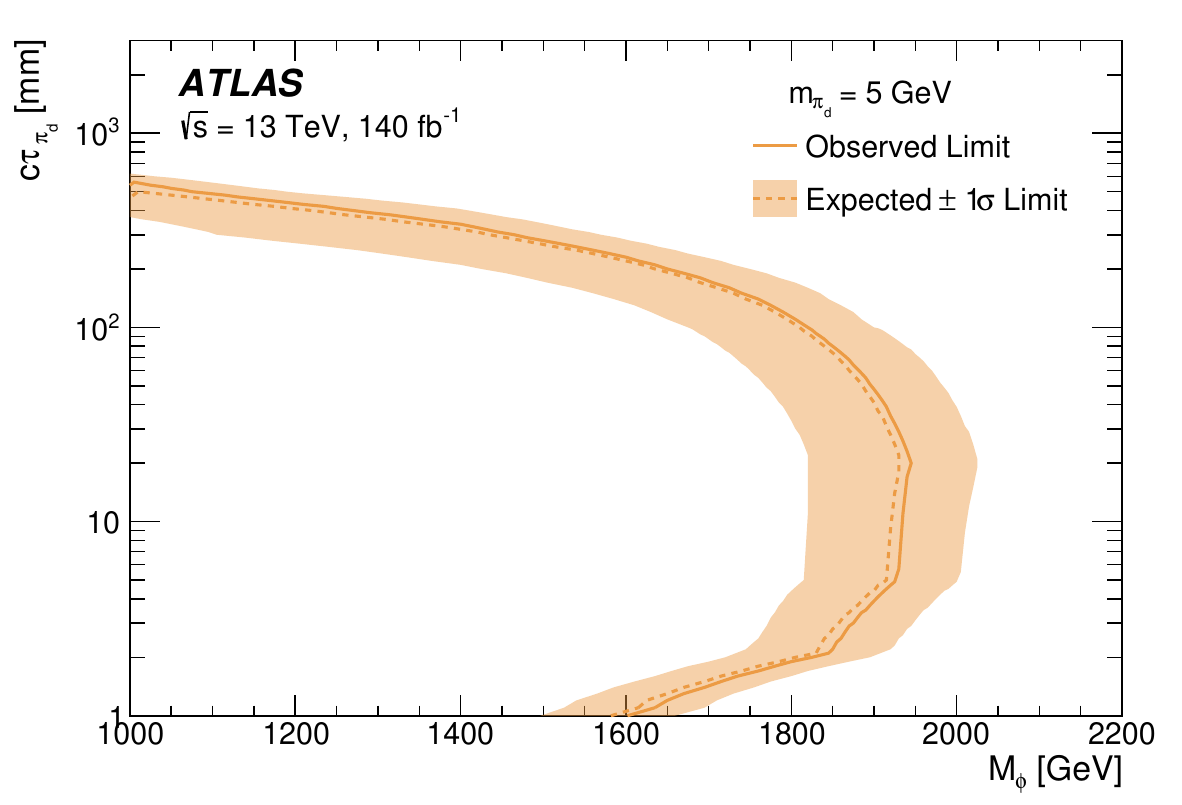}
\label{fig:CLs_mPi5}
}
\subfloat[]{
\includegraphics[width=0.48\textwidth]{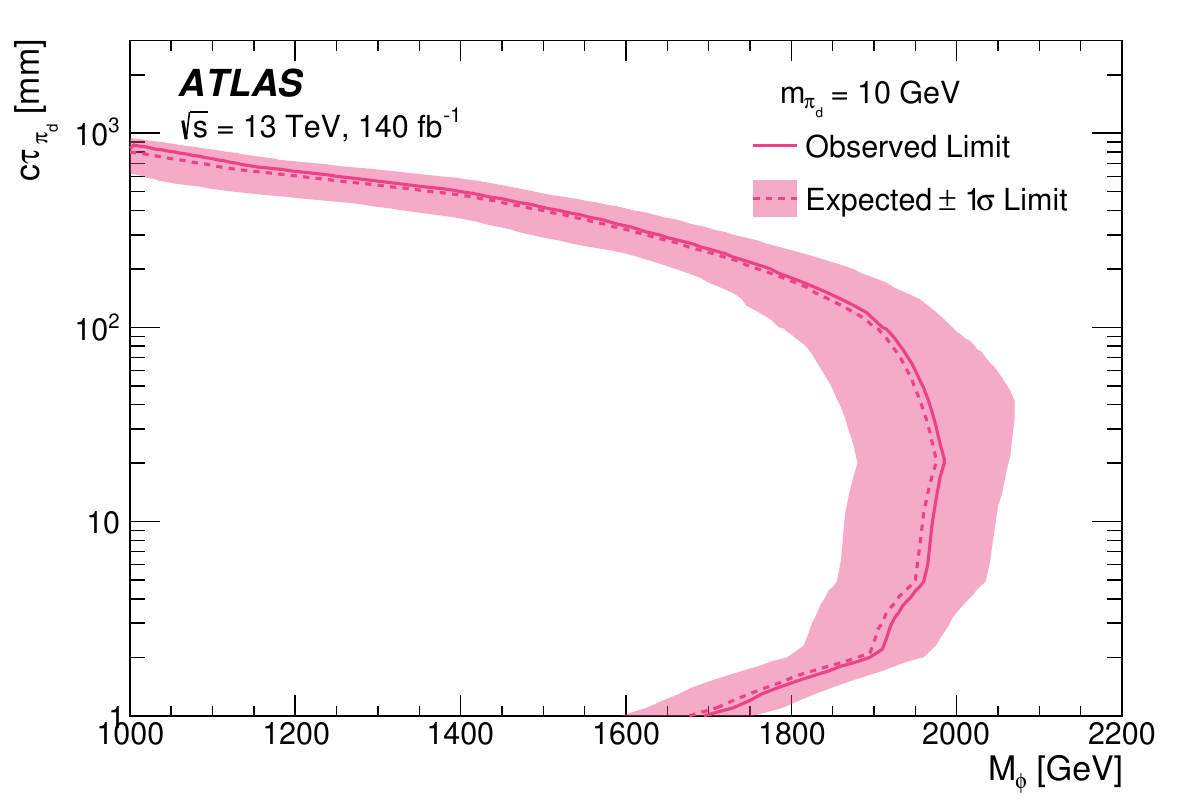}
\label{fig:CLs_mPi10}
}\\
\subfloat[]{
\includegraphics[width=0.48\textwidth]{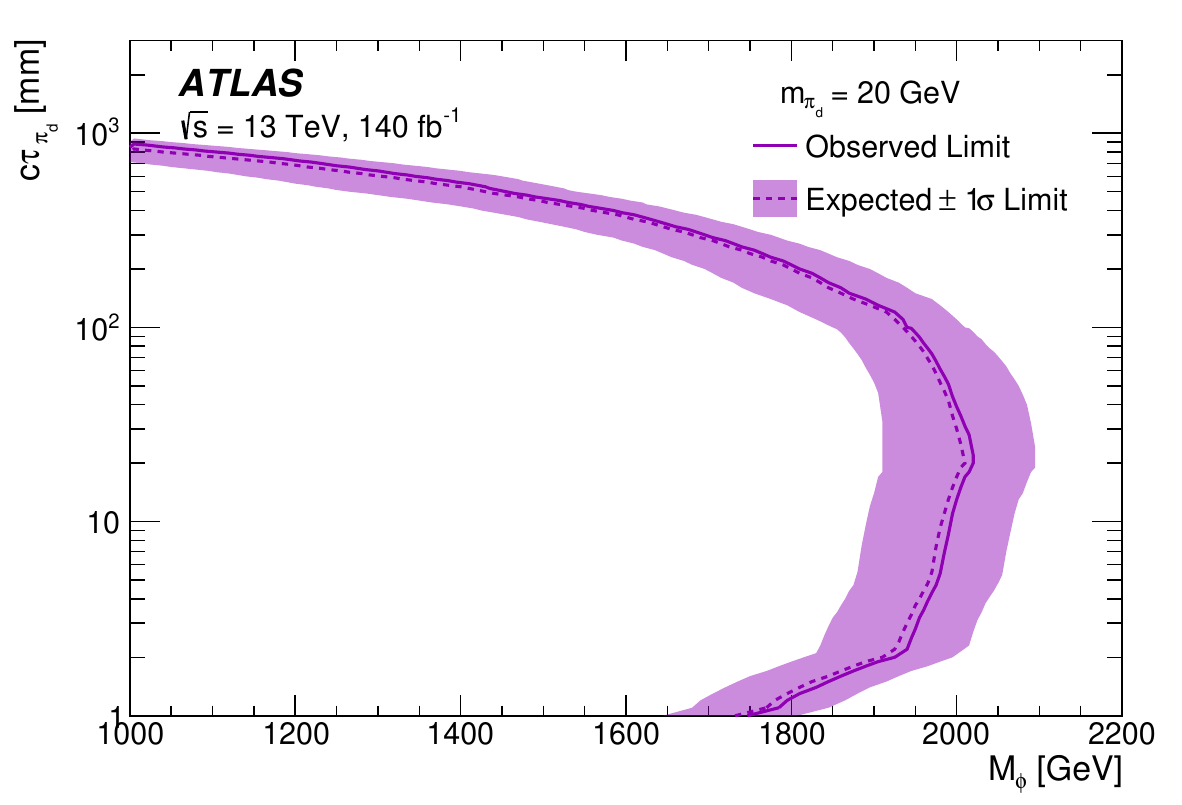}
\label{fig:CLs_mPi20}
}
\subfloat[]{
\includegraphics[width=0.48\textwidth]{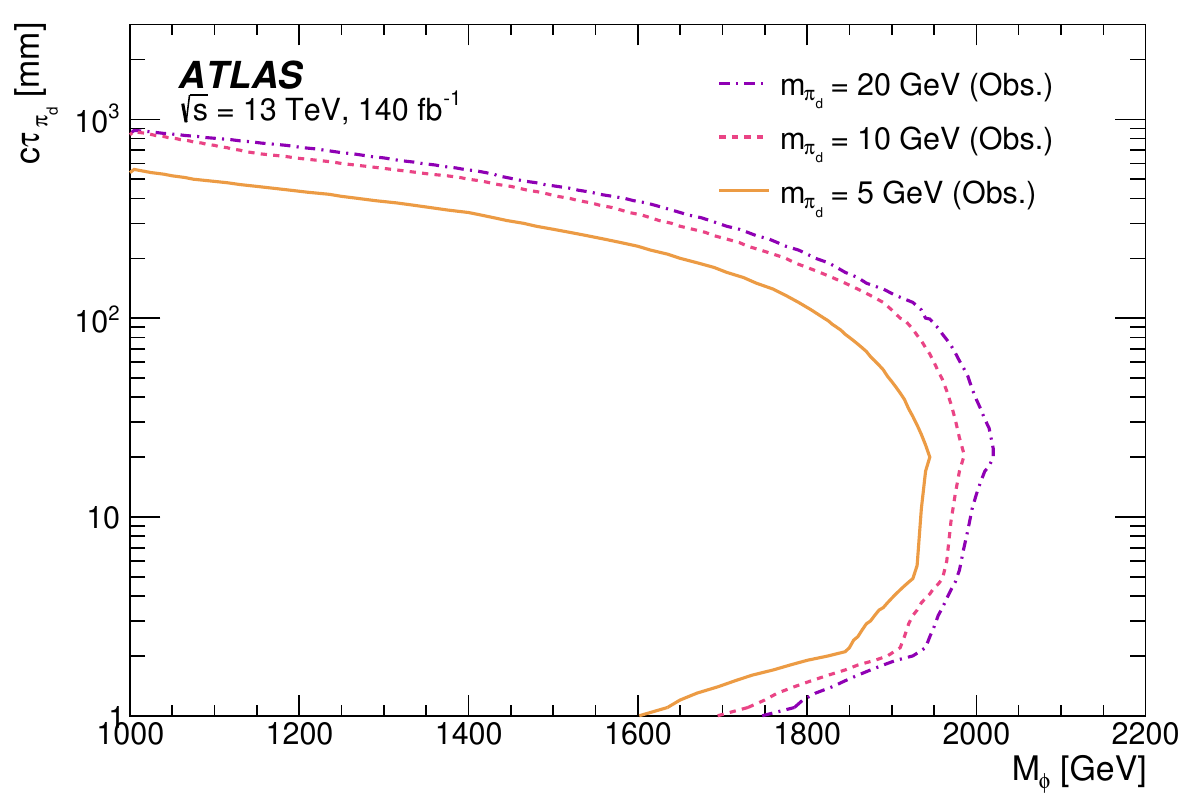}
\label{fig:Exclusion}
}
\caption{$95\%$ CL exclusion contours as a function of $M_{\phi}$ and $c\tau_{\pi_{d}}$. The expected and observed limits on $\sigma(pp \to \phi^{\dagger}\phi)$ are shown for \protect\subref{fig:CLs_mPi5} $m_{\pi_{d}} = 5~\GeV$,  \protect\subref{fig:CLs_mPi10} $m_{\pi_{d}} = 10~\GeV$, and \protect\subref{fig:CLs_mPi20} $m_{\pi_{d}} = 20~\GeV$. A comparison of the observed limits is given in \protect\subref{fig:Exclusion} for each of the three dark pion masses.}
\label{fig:CLs_mPi}
\end{figure}

Upper limits at the $95\%$ confidence level (CL) are set on the pair-production cross-section for each signal hypothesis using the $\mathrm{CL_{s}}$ method~\cite{Read:2002hq}. Figure~\ref{fig:CLs_MX_ct} shows the upper limit on $\sigma(pp \to \phi^{\dagger}\phi)$ at $95\%$ confidence level as either a function of $c\tau_{\pi_{d}}$ or $M_{\phi}$. In general, for a given combination of $M_{\phi}$ and $c\tau_{\pi_{d}}$, the sensitivity improves slightly with increasing $m_{\pi_{d}}$, as can be seen in Figures~\ref{fig:MX1400} and~\ref{fig:ct1}. The limits are strictest for intermediate dark pion lifetimes in the range $2~\mathrm{mm} < c\tau_{\pi_{d}} < 100~\mathrm{mm}$, but sensitivity decreases for $c\tau_{\pi_{d}} > 100~\mathrm{mm}$ due to the reduction in tracking and vertexing efficiency at large radii. Similarly, the limits weaken for $c\tau_{\pi_{d}} < 2~\mathrm{mm}$ due to the strict requirements on reconstructed DVs, and dark pions producing jets with larger prompt track fractions. This results in more Standard Model-like minPTF distributions, which is the strongest variable in the event-level BDT. The limits have only a weak dependence on $M_{\phi}$, getting stronger at higher mediator masses.

\begin{figure}[h]
\centering
\subfloat[]{
\includegraphics[width=0.48\textwidth]{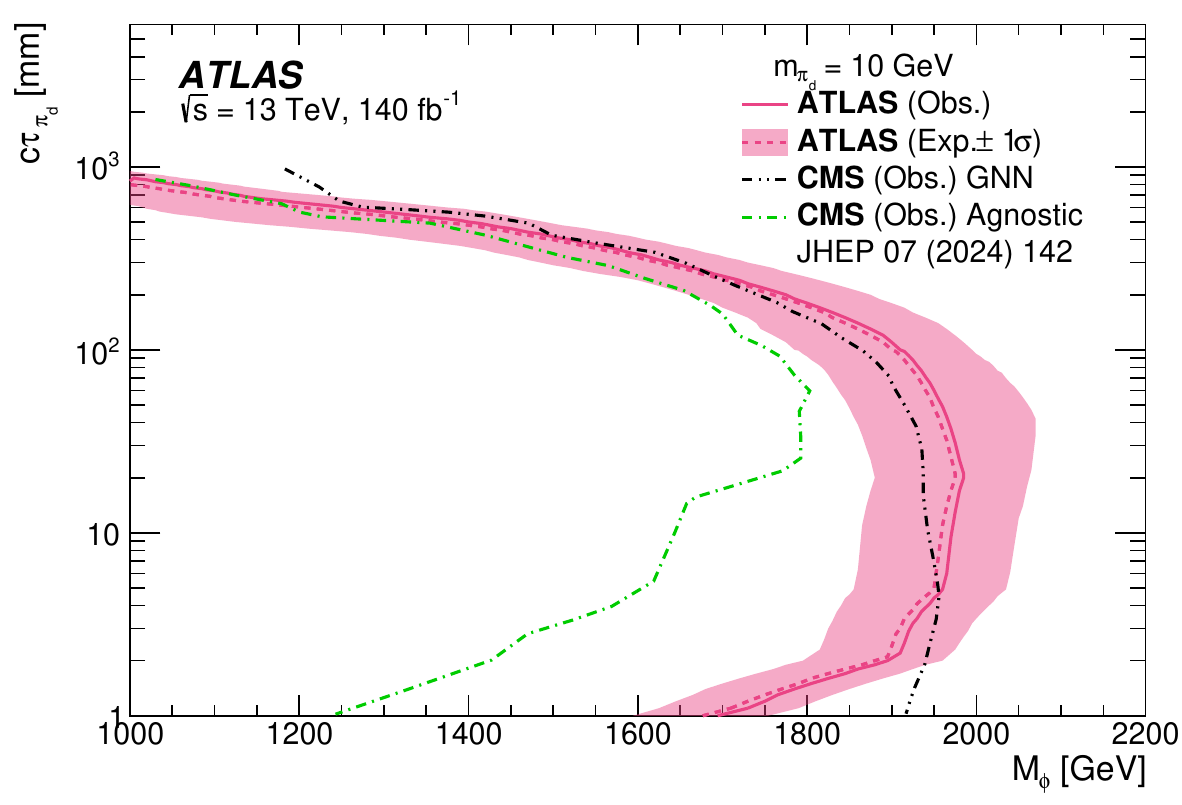}
\label{fig:mPi10_CMS}
}
\subfloat[]{
\includegraphics[width=0.48\textwidth]{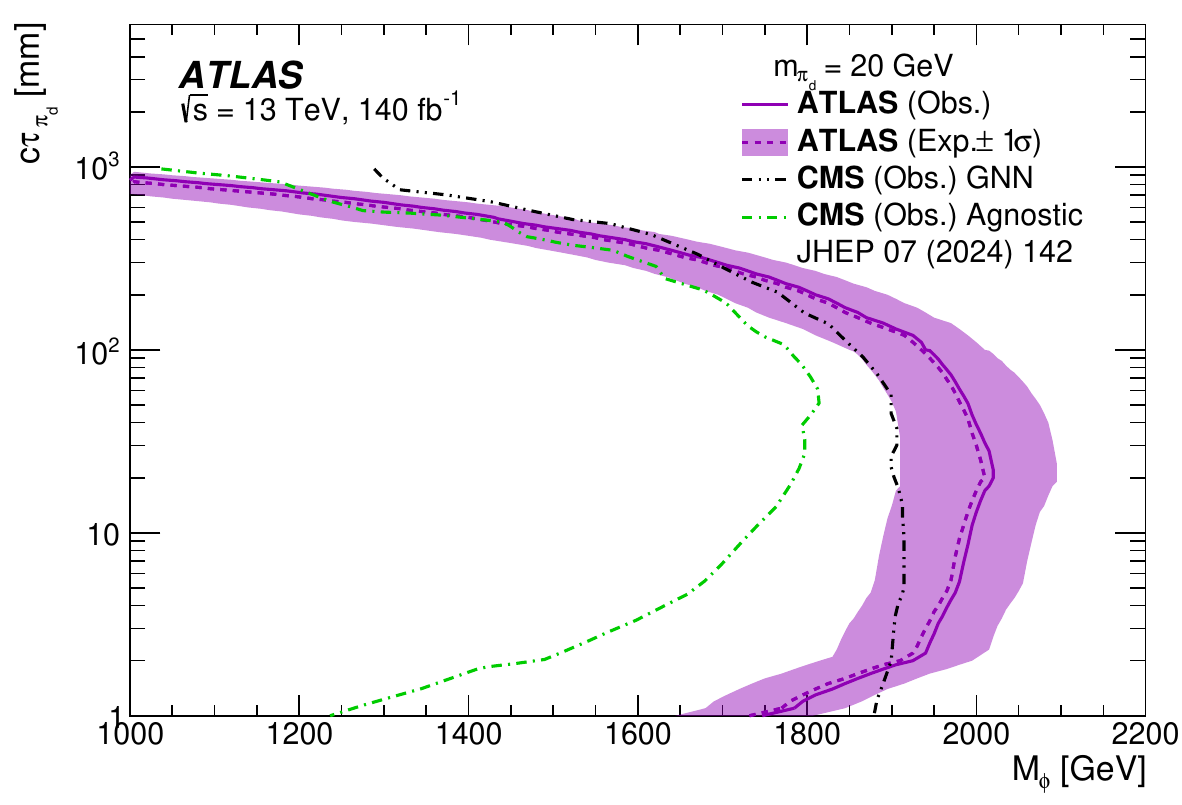}
\label{fig:mPi20_CMS}
}
\caption{Comparison of the $95\%$ CL exclusion contours between ATLAS and CMS~\cite{CMS-EXO-22-015} for \protect\subref{fig:mPi10_CMS} models with $m_{\pi_{d}} = 10~\GeV$ and \protect\subref{fig:mPi20_CMS} models with $m_{\pi_{d}} = 20~\GeV$. The expected and observed limits are shown for ATLAS, while only the observed limits in both the model agnostic and GNN cases are included for CMS. The expected limits from the CMS result are omitted, but are, in general, very close to the CMS observed limits.}
\label{fig:CLs_CMS}
\end{figure}

Exclusion contours based on the expected and observed limits on $\sigma(pp \to \phi^{\dagger}\phi)$ as a function of $M_{\phi}$ and $c\tau_{\pi_{d}}$ are shown for $m_{\pi_{d}} = 5, 10~\mathrm{and}~20~\GeV$ in Figures~\ref{fig:CLs_mPi5}, \ref{fig:CLs_mPi10}, and \ref{fig:CLs_mPi20} respectively. Figure~\ref{fig:Exclusion} gives a comparison of the exclusion contours based on the observed limits for the three dark pion masses. For models with $c\tau_{\pi_{d}} = 20~\mathrm{mm}$, the production cross-section is excluded for $\phi$ masses up to $M_{\phi} = 1.95~\TeV$, $1.99~\TeV$, and $2.00~\TeV$ for $m_{\pi_{d}} = 5,~10,~20~\GeV$, respectively. For $M_{\phi} = 1~\TeV$, dark pion lifetimes up to about $880~\mathrm{mm}$ are excluded for $m_{\pi_{d}} = 10,20~\GeV$ and up to about $530~\mathrm{mm}$ for $m_{\pi_{d}} = 5~\GeV$. The cutoff in the exclusion contours at $M_{\phi} > 2~\TeV$ is due to the rapidly falling theoretical cross-sections, which become inaccessible with the current dataset.

This analysis is able to probe cross-sections that are a factor of two smaller than those in the four jet topology of the bifundamental scalar interpretation in the ATLAS Run 3 search~\cite{EXOT-2022-31}, which is more optimised for the $Z'$ scenario.

Figure~\ref{fig:CLs_CMS} shows the comparison between the results of this analysis, with those from the corresponding CMS search~\cite{CMS-EXO-22-015}. CMS included two methods of tagging emerging jets: a \textit{model agnostic} version based on track displacement variables, and an ML-based Graph Neural Network (GNN) tagger. The GNN sets stricter limits at very low and very high proper lifetimes, while this search is more sensitive for intermediate lifetimes ($2-300~\mathrm{mm}$).

\FloatBarrier

\section{Conclusion}
\label{sec:conclusion}

This paper presents a search for emerging jets, a detector signature arising from a strongly self-interacting dark sector, using $140~\ifb$ of data from proton--proton collisions at $\sqrt{s} = 13~\TeV$, collected by the ATLAS detector at the LHC between 2015 and 2018. The search uses a two-stage boosted decision tree: the first trained on jet-level variables, and the second using event-level observables to select emerging-jet-like events from a wide model space. The Standard Model background is estimated using a fully data-driven approach. No excess above this estimate is observed.

Upper limits on the pair-production cross-section of heavy scalar mediators are set at the $95\%$ confidence level. Mediator masses up to $2~\TeV$ are excluded for dark pions with proper decay lengths of $20~\mathrm{mm}$ and a dark pion mass of $20~\GeV$. This search complements the larger experimental efforts of the ATLAS collaboration looking for dark QCD, and is competitive with the recent results from both CMS and ATLAS, in some cases setting the strictest limits to date on the pair-production of a new scalar mediator.
\section*{Acknowledgements}
%

%
%

%
%

We thank CERN for the very successful operation of the LHC and its injectors, as well as the support staff at
CERN and at our institutions worldwide without whom ATLAS could not be operated efficiently.

The crucial computing support from all WLCG partners is acknowledged gratefully, in particular from CERN, the ATLAS Tier-1 facilities at TRIUMF/SFU (Canada), NDGF (Denmark, Norway, Sweden), CC-IN2P3 (France), KIT/GridKA (Germany), INFN-CNAF (Italy), NL-T1 (Netherlands), PIC (Spain), RAL (UK) and BNL (USA), the Tier-2 facilities worldwide and large non-WLCG resource providers. Major contributors of computing resources are listed in Ref.~\cite{ATL-SOFT-PUB-2025-001}.

We gratefully acknowledge the support of ANPCyT, Argentina; YerPhI, Armenia; ARC, Australia; BMWFW and FWF, Austria; ANAS, Azerbaijan; CNPq and FAPESP, Brazil; NSERC, NRC and CFI, Canada; CERN; ANID, Chile; CAS, MOST and NSFC, China; Minciencias, Colombia; MEYS CR, Czech Republic; DNRF and DNSRC, Denmark; IN2P3-CNRS and CEA-DRF/IRFU, France; SRNSFG, Georgia; BMFTR, HGF and MPG, Germany; GSRI, Greece; RGC and Hong Kong SAR, China; ICHEP and Academy of Sciences and Humanities, Israel; INFN, Italy; MEXT and JSPS, Japan; CNRST, Morocco; NWO, Netherlands; RCN, Norway; MNiSW, Poland; FCT, Portugal; MNE/IFA, Romania; MSTDI, Serbia; MSSR, Slovakia; ARIS and MVZI, Slovenia; DSI/NRF, South Africa; MICIU/AEI, Spain; SRC and Wallenberg Foundation, Sweden; SERI, SNSF and Cantons of Bern and Geneva, Switzerland; NSTC, Taipei; TENMAK, T\"urkiye; STFC/UKRI, United Kingdom; DOE and NSF, United States of America.

Individual groups and members have received support from BCKDF, CANARIE, CRC and DRAC, Canada; CERN-CZ, FORTE and PRIMUS, Czech Republic; COST, ERC, ERDF, Horizon 2020, ICSC-NextGenerationEU and Marie Sk{\l}odowska-Curie Actions, European Union; Investissements d'Avenir Labex, Investissements d'Avenir Idex and ANR, France; DFG and AvH Foundation, Germany; Herakleitos, Thales and Aristeia programmes co-financed by EU-ESF and the Greek NSRF, Greece; BSF-NSF and MINERVA, Israel; NCN and NAWA, Poland; La Caixa Banking Foundation, CERCA Programme Generalitat de Catalunya and PROMETEO and GenT Programmes Generalitat Valenciana, Spain; G\"{o}ran Gustafssons Stiftelse, Sweden; The Royal Society and Leverhulme Trust, United Kingdom.

In addition, individual members wish to acknowledge support from CERN: European Organization for Nuclear Research (CERN DOCT); Chile: Agencia Nacional de Investigaci\'on y Desarrollo (FONDECYT 1230812, FONDECYT 1240864, Fondecyt 3240661, Fondecyt Regular 1240721); China: Chinese Ministry of Science and Technology (MOST-2023YFA1605700, MOST-2023YFA1609300), National Natural Science Foundation of China (NSFC - 12175119, NSFC 12275265); Czech Republic: Czech Science Foundation (GACR - 24-11373S), Ministry of Education Youth and Sports (ERC-CZ-LL2327, FORTE CZ.02.01.01/00/22\_008/0004632), PRIMUS Research Programme (PRIMUS/21/SCI/017); EU: H2020 European Research Council (ERC - 101002463); European Union: European Research Council (BARD No. 101116429, ERC - 948254, ERC 101089007), European Regional Development Fund (HE COFUND GA No.101081355, ERDF), European Union, Future Artificial Intelligence Research (FAIR-NextGenerationEU PE00000013), Italian Center for High Performance Computing, Big Data and Quantum Computing (ICSC, NextGenerationEU); France: Agence Nationale de la Recherche (ANR-21-CE31-0013, ANR-21-CE31-0022, ANR-22-EDIR-0002, ANR-24-CE31-0504-01); Germany: Deutsche Forschungsgemeinschaft (DFG - 469666862, DFG - CR 312/5-2); China: Research Grants Council (GRF); Italy: Istituto Nazionale di Fisica Nucleare (ICSC, NextGenerationEU), Ministero dell'Università e della Ricerca (NextGenEU 153D23001490006 M4C2.1.1, NextGenEU I53D23000820006 M4C2.1.1, NextGenEU I53D23001490006 M4C2.1.1, SOE2024\_0000023); Japan: Japan Society for the Promotion of Science (JSPS KAKENHI JP22H01227, JSPS KAKENHI JP22H04944, JSPS KAKENHI JP22KK0227, JSPS KAKENHI JP24K23939, JSPS KAKENHI JP24KK0251, JSPS KAKENHI JP25H00650, JSPS KAKENHI JP25H01291, JSPS KAKENHI JP25K01023); Norway: Research Council of Norway (RCN-314472); Poland: Ministry of Science and Higher Education (IDUB AGH, POB8, D4 no 9722), Polish National Science Centre (NCN 2021/42/E/ST2/00350, NCN OPUS 2023/51/B/ST2/02507, NCN UMO-2019/34/E/ST2/00393, UMO-2022/47/O/ST2/00148, UMO-2023/49/B/ST2/04085, UMO-2023/51/B/ST2/00920, UMO-2024/53/N/ST2/00869); Portugal: Foundation for Science and Technology (FCT); Spain: Generalitat Valenciana (ASFAE/2022/008), Ministry of Science and Innovation (MCIN \& NextGenEU PCI2022-135018-2, MICIN \& FEDER PID2021-125273NB, RYC2019-028510-I, RYC2020-030254-I, RYC2021-031273-I, RYC2022-038164-I), Ministerio de Ciencia, Innovación y Universidades/Agencia Estatal de Investigaci\'on (PID2022-142604OB-C22); Sweden: Carl Trygger Foundation (Carl Trygger Foundation CTS 22:2312), Swedish Research Council (Swedish Research Council 2023-04654, VR 2021-03651, VR 2022-03845, VR 2022-04683, VR 2023-03403, VR 2024-05451), Knut and Alice Wallenberg Foundation (KAW 2018.0458, KAW 2022.0358, KAW 2023.0366); Switzerland: Swiss National Science Foundation (SNSF - PCEFP2\_194658); United Kingdom: Royal Society (NIF-R1-231091); United States of America: U.S. Department of Energy (ECA DE-AC02-76SF00515), Neubauer Family Foundation.

%
%


%
%
%
%
%
%
%

%

\clearpage
\clearpage
\appendix
\renewcommand{\thesection}{\Alph{section}}
\setcounter{section}{0}
\section{Cut-Based Analysis}
\label{sec:appendix}

A cut-based version of the analysis was performed, but not optimized to the same degree as the BDT-based analysis. As such, it does not achieve the same sensitivity as the BDT, but it does provide results which are easier for theorists to reinterpret.

\begin{figure}[h!]
\centering
\includegraphics[width=0.6\textwidth]{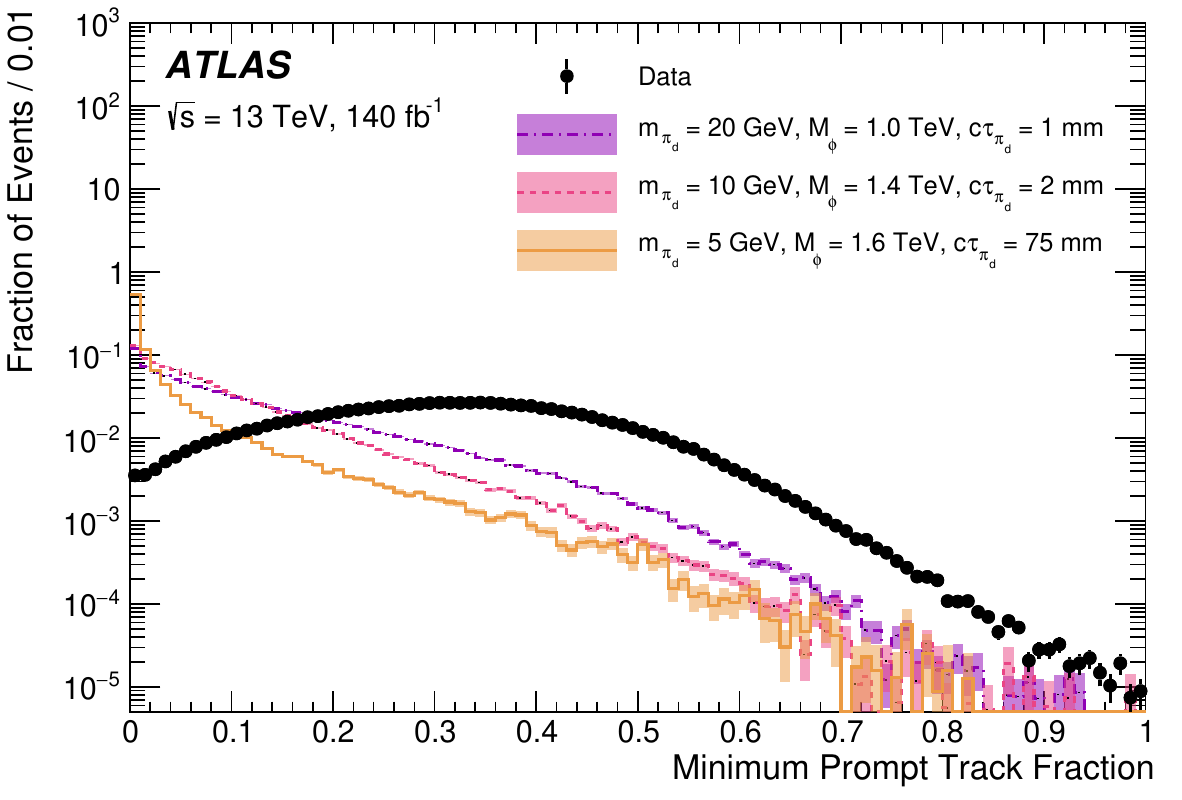}
\caption{Normalized minPTF distribution in data, with three selected signal samples for comparison}
\label{fig:minPTF}
\end{figure}

The analysis requires the same event selections as described in Section~\ref{sec:selection}, and only differs in the choice of variables used in the ABCD plane for the background estimation. Instead of using the BDT response, the cut-based analysis uses the minimum prompt track fraction of the four highest $p_{T}$ jets (minPTF), which is one of the best separating variables used in the BDT training. A comparison of the minPTF between Data and three example signal samples is shown in Figure~\ref{fig:minPTF}.

\begin{figure}[h!]
\centering
\includegraphics[width=0.6\textwidth]{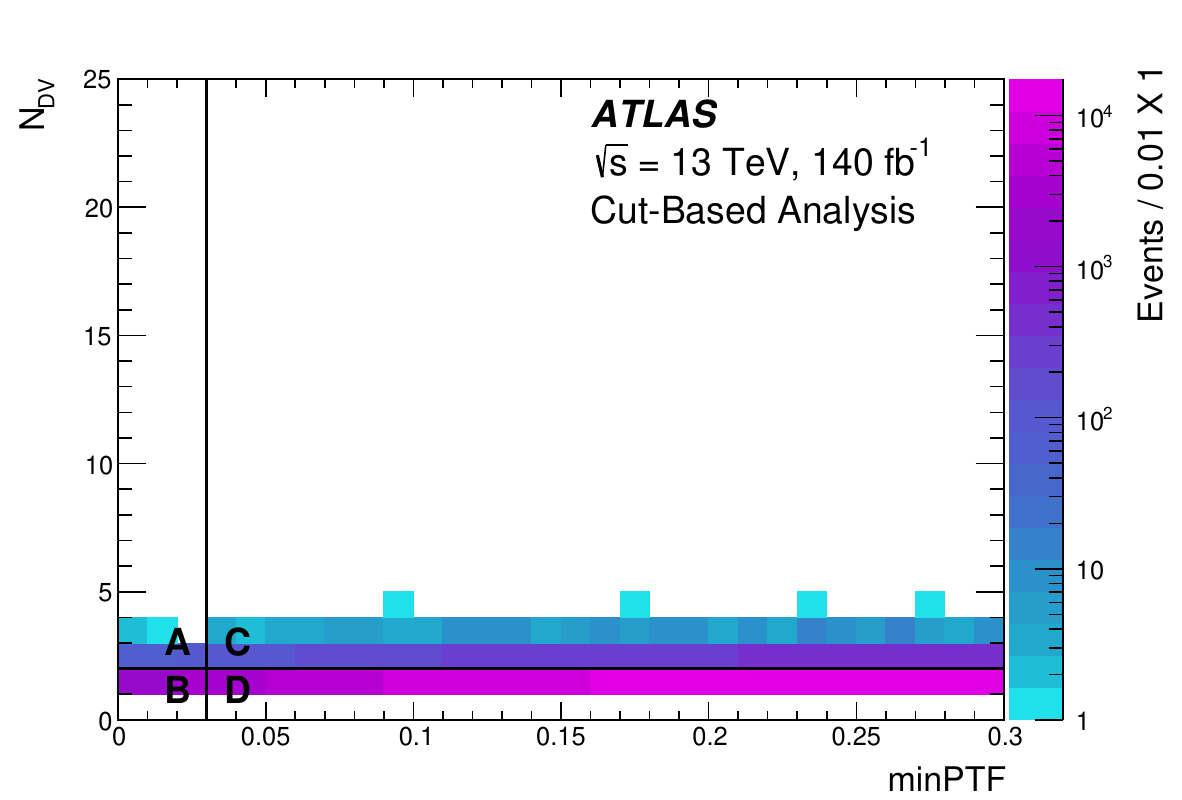}
\caption{Distribution of data events in the ABCD plane defined for the Cut-Based analysis, using the minimum prompt track fraction (minPTF) of the four highest $p_{T}$ jets on the x-axis, and the displaced vertex multiplicity on the y-axis. The cuts used to define the four regions are at minPTF $= 0.03$ and $N_{\mathrm{DV}} = 2$. The four regions are labeled accordingly.}
\label{fig:CB_ABCD}
\end{figure}

The ABCD plane is constructed using minPTF and the number of jet-matched displaced vertices in the event ($N_{\mathrm{DV}}$). The cuts used to define the signal region are minPTF $\leq 0.03$ and $N_{\mathrm{DV}} \geq 2$. In order to aid the background estimation method, which is the same as described in Section~\ref{sec:background}, minPTF is restricted to the region less than $0.3$. The distribution of data events within the ABCD plane are shown in Figure~\ref{fig:CB_ABCD}.

\begin{table}[h]
\caption{Predicted number of background events, including statistical and systematic uncertainties, and observed event yield in the Cut-Based SR.}
\label{tab:CBYield}
\centering
\begin{tabular}{ c | c  }
\hline \hline
\multicolumn{2}{c}{Cut-Based Analysis: Event Yield in the SR} \\
\hline
Prediction & $227 \pm 10~\mathrm{(stat.)} \pm 1~\mathrm{(syst.)}$ \\
Observed Yield & $219$ \\
\hline
\hline
\end{tabular}
\end{table}

An extrapolated linear fit based on the modified ABCD method was used to produce a prediction of the background contribution in the SR. The predicted and observed yield in region A are given in Table~\ref{tab:CBYield}. No excess above the SM background prediction is observed

\begin{figure}[h]
\centering
\subfloat[]{
\includegraphics[width=0.48\textwidth]{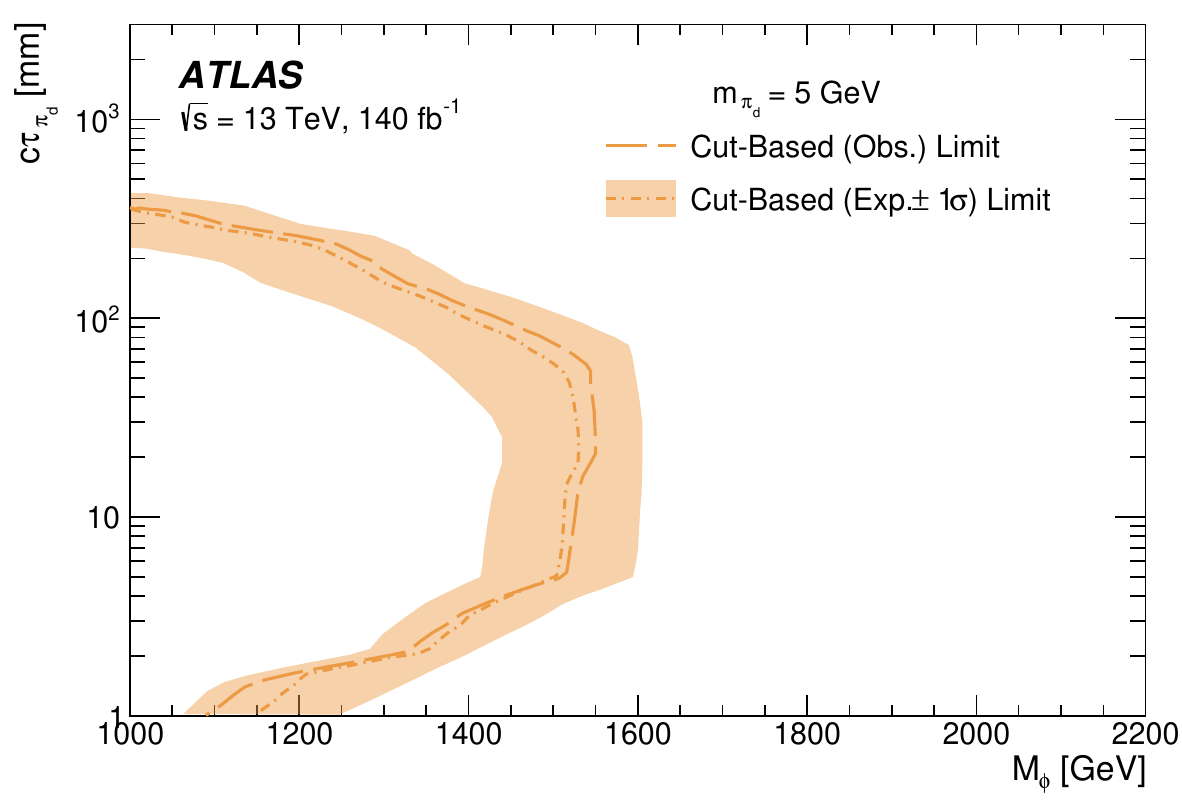}
\label{fig:CB_mPi5}
}
\subfloat[]{
\includegraphics[width=0.48\textwidth]{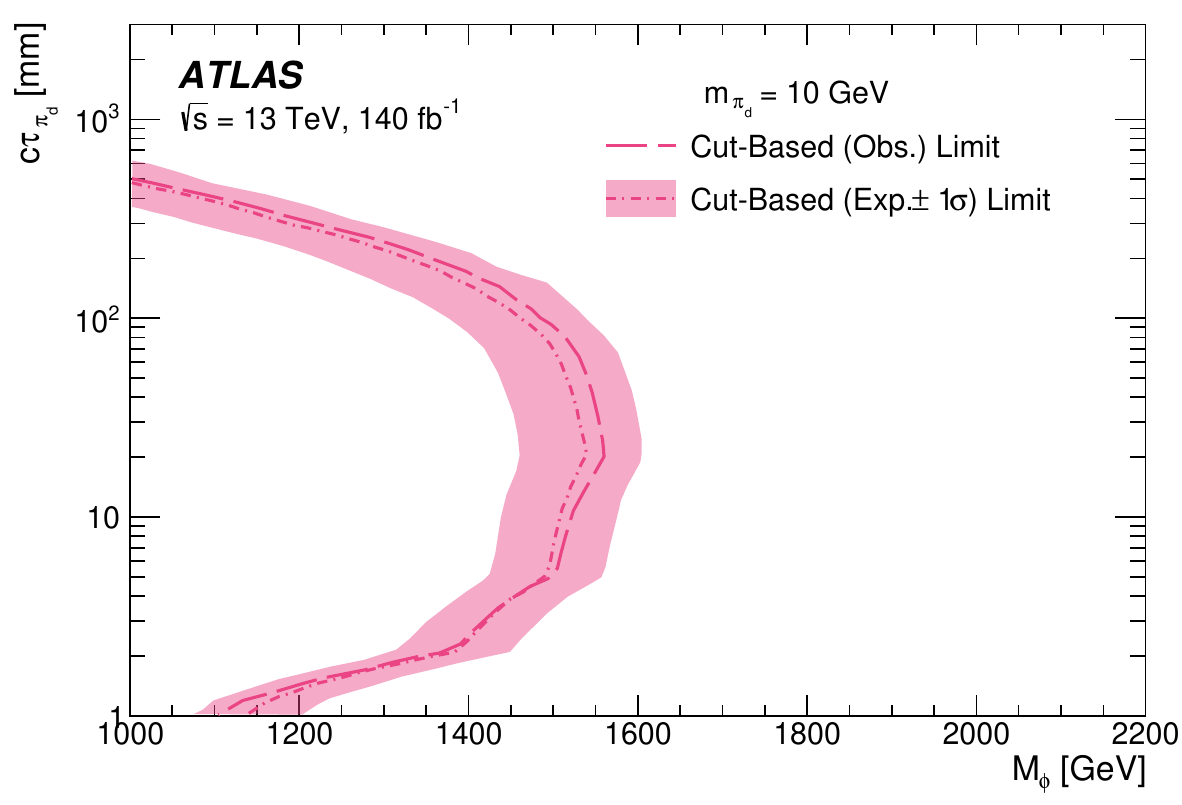}
\label{fig:CB_mPi10}
}\\
\subfloat[]{
\includegraphics[width=0.48\textwidth]{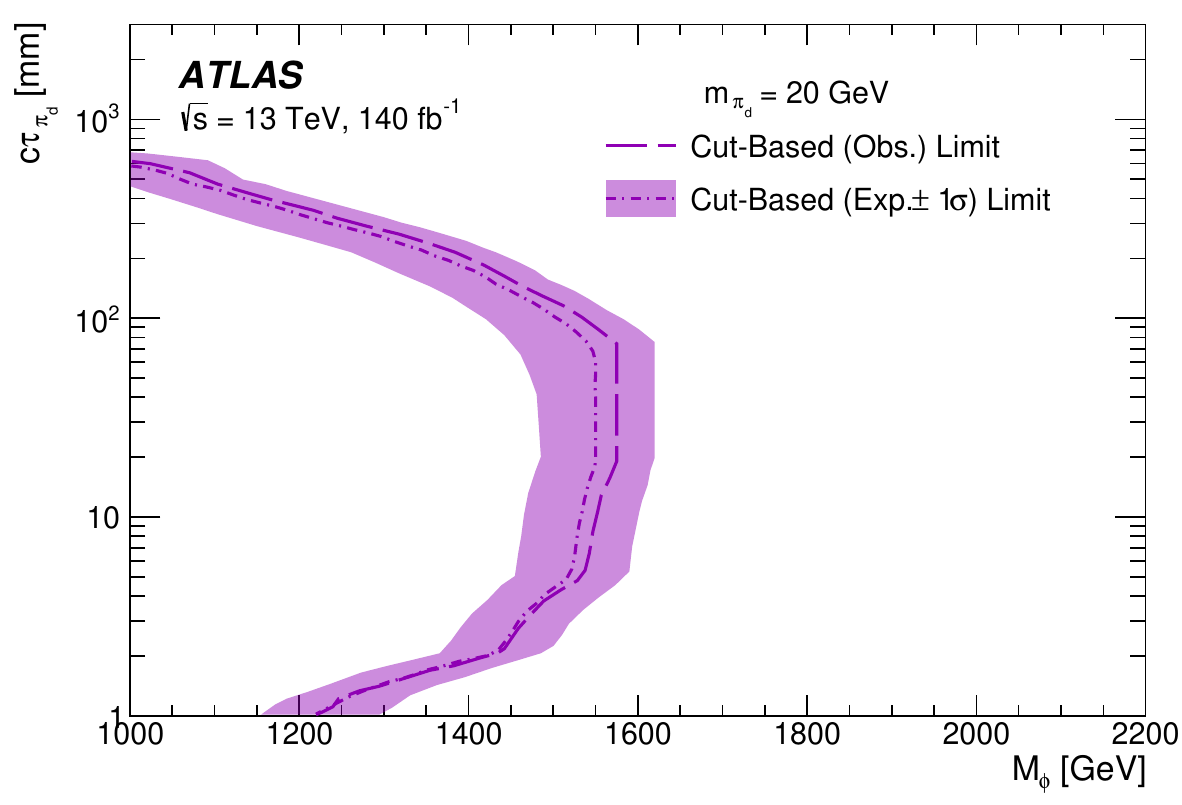}
\label{fig:CB_mPi20}
}
\subfloat[]{
\includegraphics[width=0.48\textwidth]{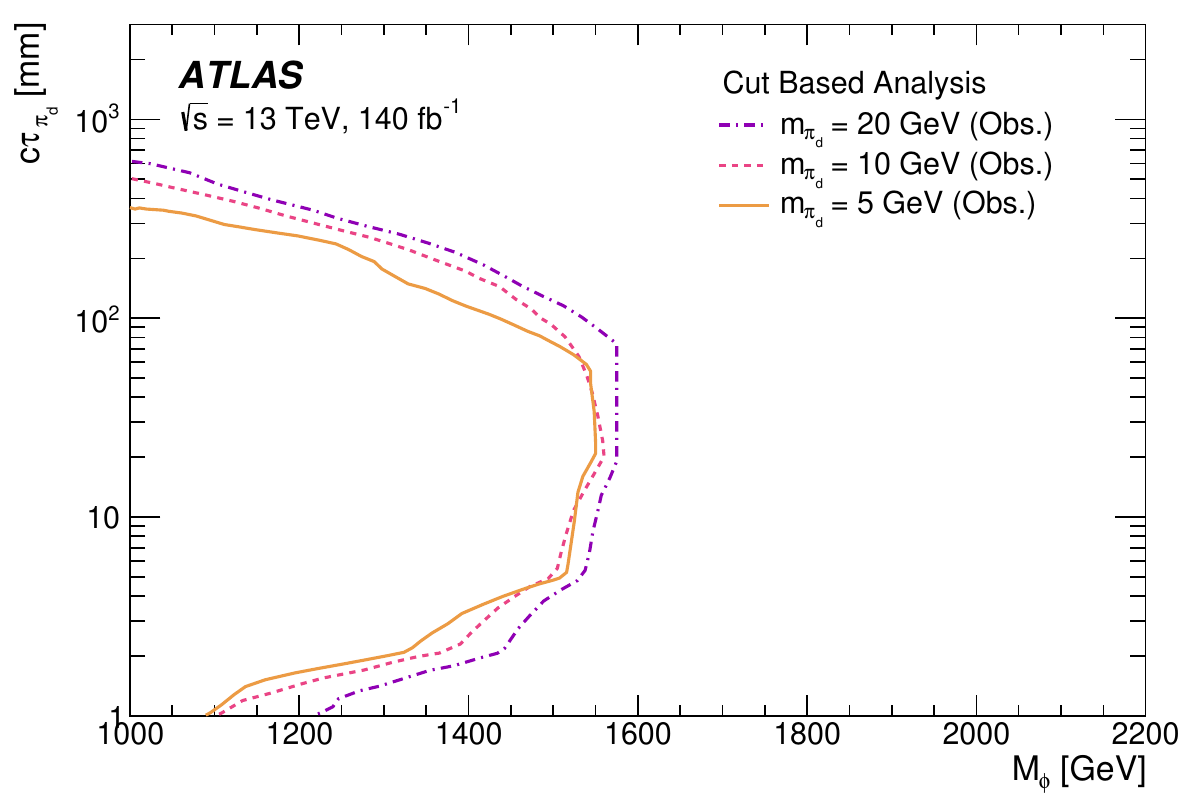}
\label{fig:CBExclusion}
}
\caption{$95\%$ CL exclusion contours as a function of $M_{\phi}$ and $c\tau_{\pi_{d}}$ for the Cut-Based analysis. The expected and observed limits on $\sigma(pp \to \phi^{\dagger}\phi)$ are shown for \protect\subref{fig:CB_mPi5} $m_{\pi_{d}} = 5~\GeV$,  \protect\subref{fig:CB_mPi10} $m_{\pi_{d}} = 10~\GeV$, and \protect\subref{fig:CB_mPi20} $m_{\pi_{d}} = 20~\GeV$. A comparison of the observed limits is given in \protect\subref{fig:CBExclusion} for each of the three dark pion masses.}
\label{fig:CB_mPi}
\end{figure}

Using the same statistical machinery as described in Section~\ref{sec:result}, limits at the $95\%$ confidence level are produced for the Cut-Based analysis. Exclusion contours as a function of $M_{\phi}$ and $c\tau_{\pi_{d}}$ are shown for $m_{\pi_{d}} = 5, 10~\mathrm{and}~20~\GeV$ in Figures~\ref{fig:CB_mPi5}, \ref{fig:CB_mPi10}, and \ref{fig:CB_mPi20} respectively. A comparison of the observed limits for all three dark pion masses is shown in Figure~\ref{fig:CBExclusion}. Similar to the BDT-based analysis, the $m_{\pi_{d}} = 20~\GeV$ models set the strictest limits of the three. Mediator masses are excluded up to $1.57~\TeV$ for $20~\GeV$ dark pions with proper lifetimes between $20~\mathrm{mm}$ and $75~\mathrm{mm}$.

\begin{figure}[h]
\centering
\subfloat[]{
\includegraphics[width=0.48\textwidth]{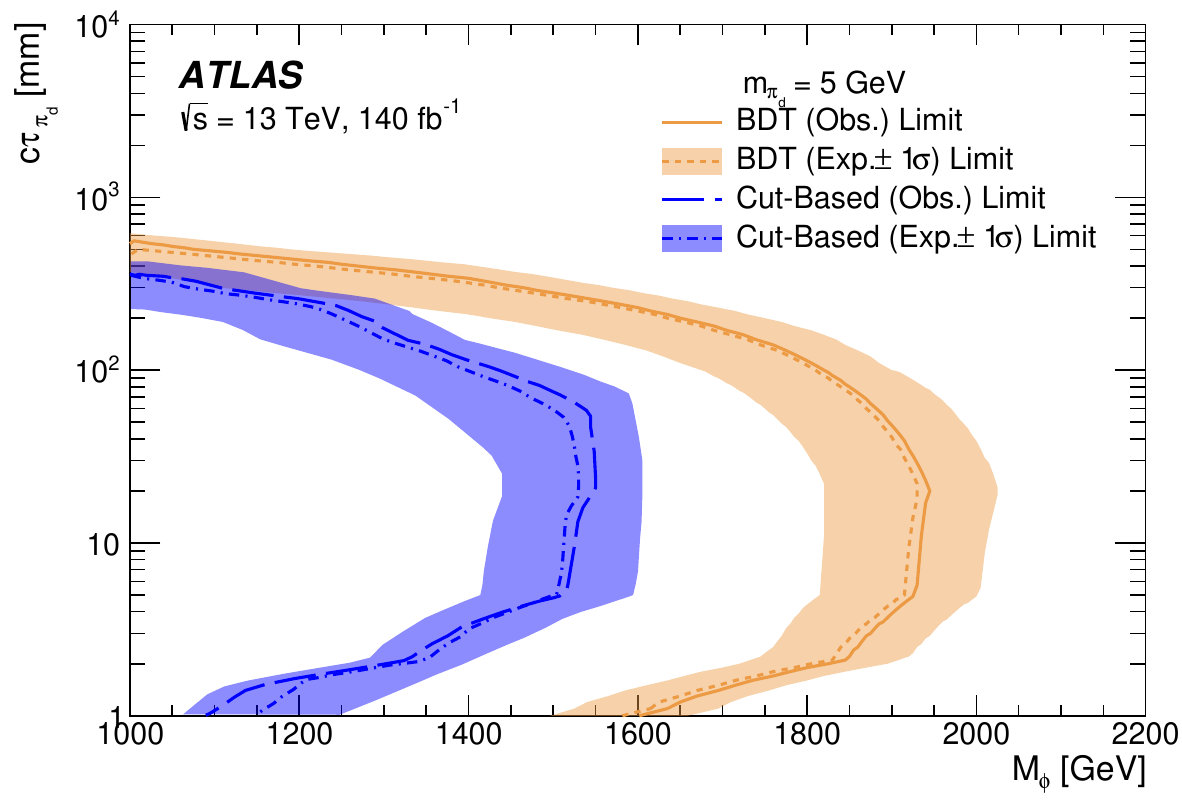}
\label{fig:Comp_CB_mPi5}
}
\subfloat[]{
\includegraphics[width=0.48\textwidth]{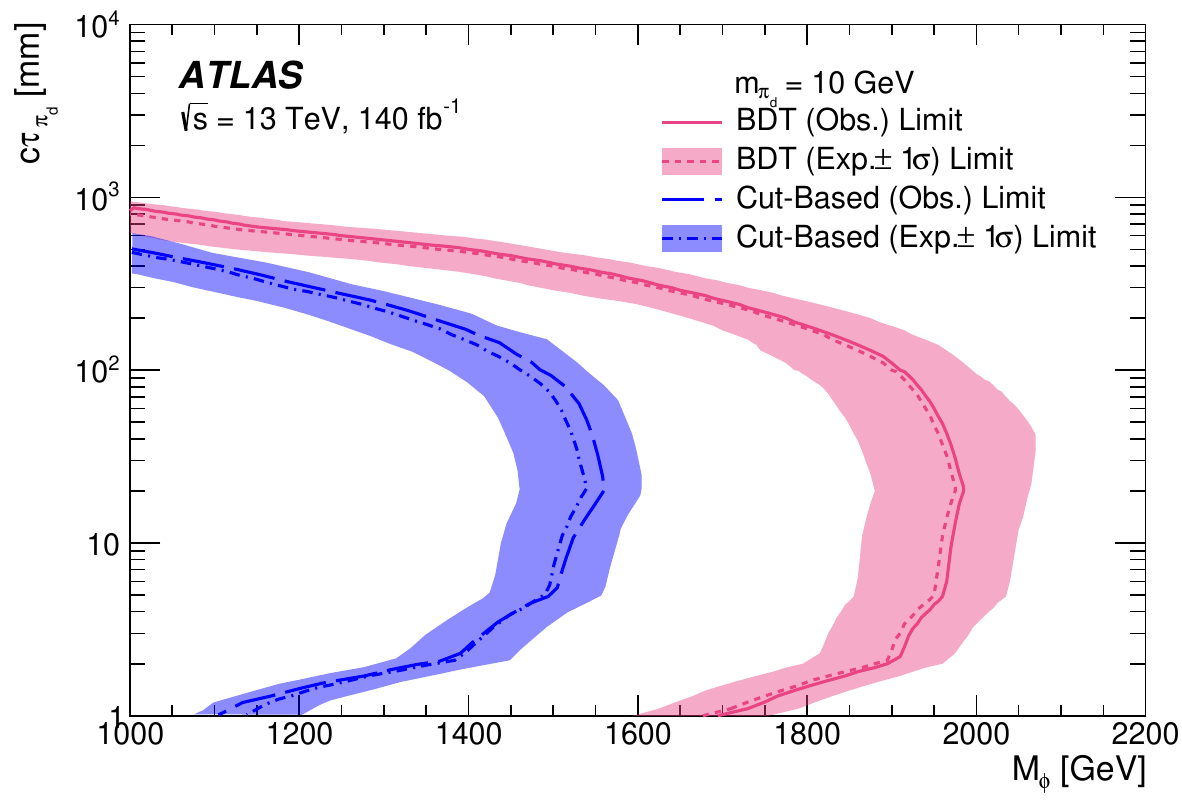}
\label{fig:Comp_CB_mPi10}
}\\
\subfloat[]{
\includegraphics[width=0.48\textwidth]{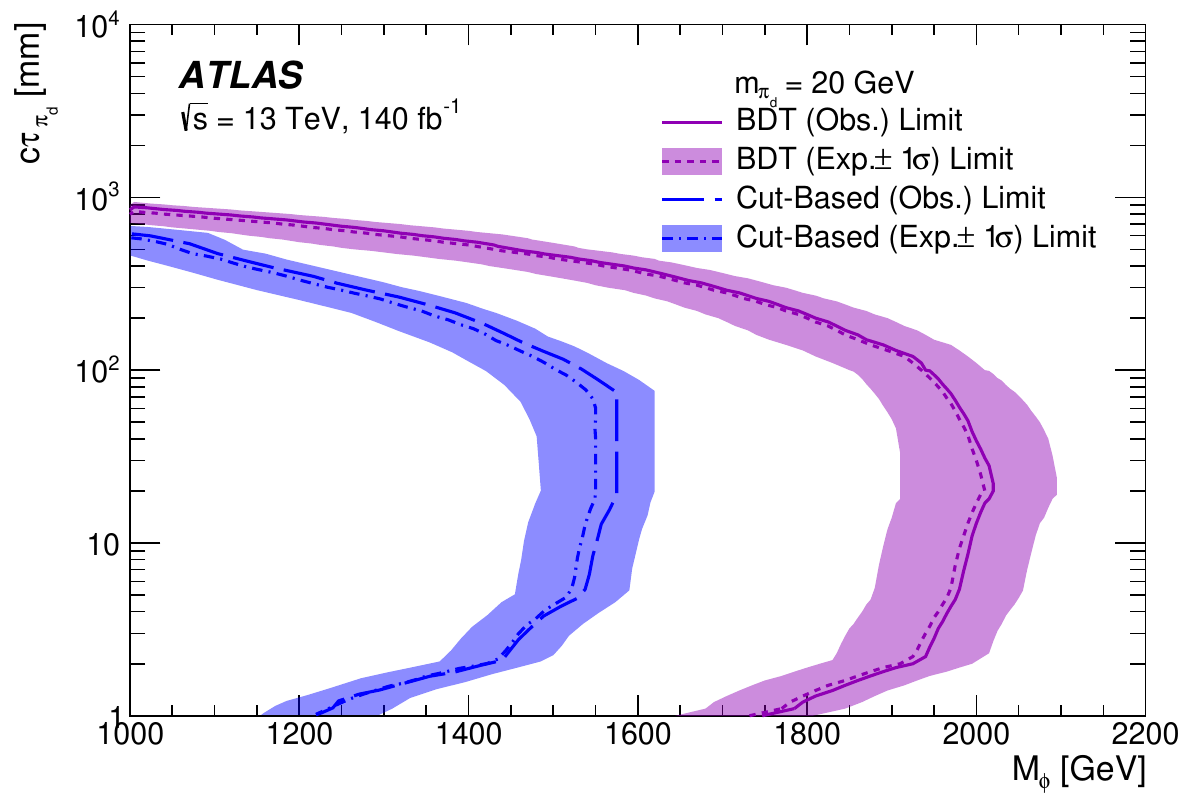}
\label{fig:Comp_CB_mPi20}
}
\caption{Comparison of observed and expected $95\%$ CL limits of the Cut-Based analysis against the corresponding BDT-based analysis. Limits on $\sigma(pp \to \phi^{\dagger}\phi)$ are shown as a function of $M_{\phi}$ and $c\tau_{\pi_{d}}$ for \protect\subref{fig:Comp_CB_mPi5} $m_{\pi_{d}} = 5~\GeV$,  \protect\subref{fig:Comp_CB_mPi10} $m_{\pi_{d}} = 10~\GeV$, and \protect\subref{fig:Comp_CB_mPi20} $m_{\pi_{d}} = 20~\GeV$.}
\label{fig:Comp_CB_mPi}
\end{figure}

The Cut-Based analysis excludes a much smaller region of phase-space as compared to the BDT-based analysis, which speaks to the enhanced separating power that the BDT response has over any of the individual training variables. Comparisons between the Cut-Based and BDT-based analyses are shown in Figure~\ref{fig:Comp_CB_mPi} for each of the dark pion masses individually.

\clearpage

\printbibliography

@article{NNLL-Fast,
   title={NNLL-fast: predictions for coloured supersymmetric particle production at the LHC with threshold and Coulomb resummation},
   volume={12},
   % ISSN={1029-8479},
   % % url={http://dx.doi.org/10.1007/JHEP12(2016)133},
   DOI={10.1007/jhep12(2016)133},
   pages={133},
   journal={JHEP},
   publisher={Springer Science and Business Media LLC},
   author={Beenakker, Wim and Borschensky, Christoph and Krämer, Michael and Kulesza, Anna and Laenen, Eric},
   year={2016},
   month=dec, 
   eprint         = "1607.07741",
   archivePrefix  = "arXiv",
   primaryClass   = "hep-ph"
}

@article{NNLLresum,
   title={NNLL resummation for stop pair-production at the LHC},
   volume={05},
   % ISSN={1029-8479},
   % % url={http://dx.doi.org/10.1007/JHEP05(2016)153},
   DOI={10.1007/jhep05(2016)153},
   pages={153},
   journal={JHEP},
   publisher={Springer Science and Business Media LLC},
   author={Beenakker, Wim and Borschensky, Christoph and Heger, Raphael and Krämer, Michael and Kulesza, Anna and Laenen, Eric},
   year={2016},
   month=may,
   eprint         = "1601.02954",
   archivePrefix  = "arXiv",
   primaryClass   = "hep-ph" 
}

@article{Cowan_2011,
   title={Asymptotic formulae for likelihood-based tests of new physics},
   volume={71},
   % ISSN={1434-6052},
   % % url={http://dx.doi.org/10.1140/epjc/s10052-011-1554-0},
   DOI={10.1140/epjc/s10052-011-1554-0},
   number={2},
   journal={Eur. Phys. J. C},
   publisher={Springer Science and Business Media LLC},
   author={Cowan, Glen and Cranmer, Kyle and Gross, Eilam and Vitells, Ofer},
   year={2011},
   month=feb,
   eprint         = "1007.1727",
   archivePrefix  = "arXiv",
   primaryClass   = "physics.data-an"  
}

@article{darkJets,
   title={Tagging a jet from a dark sector with jet substructures at colliders},
   volume={100},
   % ISSN={2470-0029},
   % url={http://dx.doi.org/10.1103/PhysRevD.100.115009},
   DOI={10.1103/physrevd.100.115009},
   number={11},
   journal={Phys. Rev. D},
   publisher={American Physical Society (APS)},
   author={Park, Myeonghun and Zhang, Mengchao},
   year={2019},
   month=dec,
   eprint         = "1712.09279",
   archivePrefix  = "arXiv",
   primaryClass   = "hep-ph"  
}

@software{xRooFit,
    author      = {Buttinger, Will},
    title       = "{Extra tools for RooFit projects}",
    url         = "https://gitlab.cern.ch/will/xroofit/-/blob/master/",
    year        = 2023
}

@article{Rubin_1980,
  title = "Rotational properties of 21 SC galaxies with a large range of luminosities and radii, from NGC 4605 ($R = 4$ kpc) to UGC 2885 ($R = 122$ kpc)",
  author = "Rubin, V. C. and Ford, Jr., W. K. and Thonnard, N.",
  journal = {ApJ},
  volume = "238",
  pages = "471",
  year = "1980",
  doi = "10.1086/158003"
}

@article{Persic_1996,
   author = {Persic, Massimo and Salucci, Paolo and Stel, Fulvio},
   title = "{The Universal Rotation Curve of Spiral Galaxies - I. The Dark Matter Connection}",
   journal = {MNRAS},
   year = {1996},
   volume = {281},
   number = {1},
   pages = {27-47},
   doi = {10.48550/arXiv.astro-ph/9506004},
   eprint         = "astro-ph/9506004",
   archivePrefix  = "arXiv",
   primaryClass   = "astro-ph" 
}

@article{Clowe_2006,
   title={A Direct Empirical Proof of the Existence of Dark Matter},
   volume={648},
   % ISSN={1538-4357},
   % url={http://dx.doi.org/10.1086/508162},
   DOI={10.1086/508162},
   number={2},
   journal={ApJ},
   publisher={American Astronomical Society},
   author={Clowe, Douglas and Bradač, Maruša and Gonzalez, Anthony H. and Markevitch, Maxim and Randall, Scott W. and Jones, Christine and Zaritsky, Dennis},
   year={2006},
   month=aug, 
   pages={L109-L113},
   eprint         = "astro-ph/0608407",
   archivePrefix  = "arXiv",
   primaryClass   = "astro-ph" 
}

@article{PandaX,
   title={Dark Matter Search Results from 1.54 Tonne $\cdot$ Year Exposure of PandaX-4T},
   volume={134},
   % ISSN={1079-7114},
   % url={http://dx.doi.org/10.1103/PhysRevLett.134.011805},
   DOI={10.1103/physrevlett.134.011805},
   pages={011805},
   journal={Phys. Rev. Lett.},
   publisher={American Physical Society (APS)},
   author={PandaX~Collaboration},
   year={2025},
   month=jan,
   eprint         = "2408.00664",
   archivePrefix  = "arXiv",
   primaryClass   = "hep-ex" 
}

@article{PICO,
   title={Dark matter Search Results from the Complete Exposure of the PICO-60 C$_{3}$F$_{8}$ Bubble Chamber},
   volume={100},
   % ISSN={2470-0029},
   % url={http://dx.doi.org/10.1103/PhysRevD.100.022001},
   DOI={10.1103/physrevd.100.022001},
   pages={022001},
   journal={Phys. Rev. D},
   publisher={American Physical Society (APS)},
   author={PICO~Collaboration},
   year={2019},
   month=jul,
   eprint         = "1902.04031",
   archivePrefix  = "arXiv",
   primaryClass   = "astro-ph" 
}

@article{DarkSide,
   title={DarkSide-50 532-day dark matter search with low-radioactivity argon},
   volume={98},
   % ISSN={2470-0029},
   % url={http://dx.doi.org/10.1103/PhysRevD.98.102006},
   DOI={10.1103/physrevd.98.102006},
   pages={102006},
   journal={Phys. Rev. D},
   publisher={American Physical Society (APS)},
   author={DarkSide~Collaboration},
   year={2018},
   month=nov,
   eprint         = "1802.07198",
   archivePrefix  = "arXiv",
   primaryClass   = "astro-ph"   
}

@article{LZ,
   title={Dark Matter Search Results from 4.2 Tonne-Years of Exposure of the LUX-ZEPLIN (LZ) Experiment},
   volume={135},
   % ISSN={1079-7114},
   % url={http://dx.doi.org/10.1103/4dyc-z8zf},
   DOI={10.1103/4dyc-z8zf},
   pages={011802},
   journal={Phys. Rev. Lett.},
   publisher={American Physical Society (APS)},
   author={LZ~Collaboration},
   year={2025},
   month=jul,
   eprint         = "2410.17036",
   archivePrefix  = "arXiv",
   primaryClass   = "hep-ex"   
}

@article{XENON,
   title={Search for Coherent Elastic Scattering of Solar $^{8}$B Neutrinos in the XENON1T Dark Matter Experiment},
   volume={126},
   % ISSN={1079-7114},
   % url={http://dx.doi.org/10.1103/PhysRevLett.126.091301},
   DOI={10.1103/physrevlett.126.091301},
   pages={091301},
   journal={Phys. Rev. Lett.},
   publisher={American Physical Society (APS)},
   author={XENON~Collaboration},
   year={2021},
   month=mar,
   eprint         = "2012.02846",
   archivePrefix  = "arXiv",
   primaryClass   = "hep-ex"  
}

@article{DEAP-3600,
   title={Recent results from DEAP-3600},
   volume={18},
   % ISSN={1748-0221},
   % url={http://dx.doi.org/10.1088/1748-0221/18/02/C02046},
   DOI={10.1088/1748-0221/18/02/c02046},
   number={02},
   journal={JINST},
   publisher={IOP Publishing},
   author={M.~Lai~(on~behalf~of~the~DEAP-3600~Collaboration)},
   year={2023},
   month=feb,
   pages={C02046},
   eprint         = "2302.14484",
   archivePrefix  = "arXiv",
   primaryClass   = "hep-ex"  
}

@article{SuperCDMS,
   title={Search for low-mass dark matter with CDMSlite using a profile likelihood fit},
   volume={99},
   % ISSN={2470-0029},
   % url={http://dx.doi.org/10.1103/PhysRevD.99.062001},
   DOI={10.1103/physrevd.99.062001},
   pages={062001},
   journal={Phys. Rev. D},
   publisher={American Physical Society (APS)},
   author={SuperCDMS~Collaboration},
   year={2019},
   month=mar,
   eprint         = "1808.09098",
   archivePrefix  = "arXiv",
   primaryClass   = "astro-ph" 
}

@article{Strassler_2007,
   title={Echoes of a hidden valley at hadron colliders},
   volume={651},
   % ISSN={0370-2693},
   % url={http://dx.doi.org/10.1016/j.physletb.2007.06.055},
   DOI={10.1016/j.physletb.2007.06.055},
   number={5–6},
   journal={Phys. Lett. B},
   publisher={Elsevier BV},
   author={Strassler, Matthew J. and Zurek, Kathryn M.},
   year={2007},
   month=aug, 
   pages={374},
   eprint         = "hep-ph/0604261",
   archivePrefix  = "arXiv",
   primaryClass   = "hep-ph" 
}

@article{Strassler_2008,
   title={Discovering the Higgs through highly-displaced vertices},
   volume={661},
   % ISSN={0370-2693},
   % url={http://dx.doi.org/10.1016/j.physletb.2008.02.008},
   DOI={10.1016/j.physletb.2008.02.008},
   number={4},
   journal={Phys. Lett. B},
   publisher={Elsevier BV},
   author={Strassler, Matthew J. and Zurek, Kathryn M.},
   year={2008},
   month=mar, 
   pages={263-267},
   eprint         = "hep-ph/0605193",
   archivePrefix  = "arXiv",
   primaryClass   = "hep-ph" 
}

@article{Han_2008,
   title={Phenomenology of hidden valleys at hadron colliders},
   volume={07},
   % ISSN={1029-8479},
   % url={http://dx.doi.org/10.1088/1126-6708/2008/07/008},
   DOI={10.1088/1126-6708/2008/07/008},
   number={008},
   journal={JHEP},
   publisher={Springer Science and Business Media LLC},
   author={Han, Tao and Si, Zongguo and Zurek, Kathryn M and Strassler, Matthew J},
   year={2008},
   month=jul, 
   pages={008},
   eprint         = "0712.2041",
   archivePrefix  = "arXiv",
   primaryClass   = "hep-ph" 
}

@article{SSW,
   title={Emerging jets},
   volume={05},
   % ISSN={1029-8479},
   % url={http://dx.doi.org/10.1007/JHEP05(2015)059},
   DOI={10.1007/jhep05(2015)059},
   pages={059},
   journal={JHEP},
   publisher={Springer Science and Business Media LLC},
   author={Schwaller, Pedro and Stolarski, Daniel and Weiler, Andreas},
   year={2015},
   month=may,
   eprint         = "1502.05409",
   archivePrefix  = "arXiv",
   primaryClass   = "hep-ph"  
}

@article{Bai_2014,
   title={Scale of dark QCD},
   volume={89},
   % ISSN={1550-2368},
   % url={http://dx.doi.org/10.1103/PhysRevD.89.063522},
   DOI={10.1103/physrevd.89.063522},
   pages={063522},
   journal={Phys. Rev. D},
   publisher={American Physical Society (APS)},
   author={Bai, Yang and Schwaller, Pedro},
   year={2014},
   month=mar,
   eprint         = "1306.4676",
   archivePrefix  = "arXiv",
   primaryClass   = "hep-ph"  
}

@article{Beauchesne_2019,
   title={Dark matter in Hidden Valley models with stable and unstable light dark mesons},
   volume={04},
   % ISSN={1029-8479},
   % url={http://dx.doi.org/10.1007/JHEP04(2019)118},
   DOI={10.1007/jhep04(2019)118},
   pages={118},
   journal={JHEP},
   publisher={Springer Science and Business Media LLC},
   author={Beauchesne, Hugues and Bertuzzo, Enrico and Grilli di Cortona, Giovanni},
   year={2019},
   month=apr,
   eprint         = "1809.10152",
   archivePrefix  = "arXiv",
   primaryClass   = "hep-ph"  
}

@article{Albouy_2022,
   title={Theory, phenomenology, and experimental avenues for dark showers: a Snowmass 2021 report},
   volume={82},
   % ISSN={1434-6052},
   % url={http://dx.doi.org/10.1140/epjc/s10052-022-11048-8},
   DOI={10.1140/epjc/s10052-022-11048-8},
   pages={1132},
   journal={Eur. Phys. J. C},
   publisher={Springer Science and Business Media LLC},
   author={Albouy, Guillaume and Barron, Jared and Beauchesne, Hugues and Bernreuther, Elias and Bona, Marcella and Cazzaniga, Cesare and Cesarotti, Cari and Cohen, Timothy and de Cosa, Annapaola and Curtin, David and Demiragli, Zeynep and Doglioni, Caterina and Elliot, Alison and DiPetrillo, Karri Folan and Eble, Florian and Erice, Carlos and Freer, Chad and Garcia-Bellido, Aran and Gemmell, Caleb and Genest, Marie-Hélène and di Cortona, Giovanni Grilli and Gustavino, Giuliano and Hemme, Nicoline and Holmes, Tova and Kar, Deepak and Knapen, Simon and Kulkarni, Suchita and Lavezzo, Luca and Lowette, Steven and Maier, Benedikt and Mee, Seán and Mrenna, Stephen and Nair, Harikrishnan and Niedziela, Jeremi and Papageorgakis, Christos and Parmar, Nukulsinh and Paus, Christoph and Pedro, Kevin and Peixoto, Ana and Perloff, Alexx and Plehn, Tilman and Scherb, Christiane and Schwaller, Pedro and Shelton, Jessie and Singh, Akanksha and Sinha, Sukanya and Sjöstrand, Torbjörn and Spourdalakis, Aris G. B. and Stolarski, Daniel and Strassler, Matthew J. and Usachov, Andrii and Sierra, Carlos Vázquez and Verhaaren, Christopher B. and Wang, Long},
   year={2022},
   month=dec,
   eprint         = "2203.09503",
   archivePrefix  = "arXiv",
   primaryClass   = "hep-ph"  
}

@article{Cohen_2015,
   title={Semivisible Jets: Dark Matter Undercover at the LHC},
   volume={115},
   % ISSN={1079-7114},
   % url={http://dx.doi.org/10.1103/PhysRevLett.115.171804},
   DOI={10.1103/physrevlett.115.171804},
   number={17},
   journal={Phys. Rev. Lett.},
   publisher={American Physical Society (APS)},
   author={Cohen, Timothy and Lisanti, Mariangela and Lou, Hou Keong},
   year={2015},
   month=oct,
   eprint         = "1503.00009",
   archivePrefix  = "arXiv",
   primaryClass   = "hep-ph" 
}

@article{tChannel,
      title={t-channel dark matter models - a whitepaper},
      volume={85},
      % ISSN={1434-6052},
      % url={http://dx.doi.org/10.1140/epjc/s10052-025-14635-7},
      DOI={10.1140/epjc/s10052-025-14635-7},
      pages={975},
      journal={Eur. Phys. J. C},
      publisher={Springer Science and Business Media LLC},
      author={Arina, Chiara and Fuks, Benjamin and Panizzi, Luca and Baker, Michael J. and Cornell, Alan S. and Heisig, Jan and Maier, Benedikt and Pedro, Rute and Trischuk, Dominique and Agin, Diyar and Arbey, Alexandre and Arcadi, Giorgio and Bagnaschi, Emanuele and Bai, Kehang and Bhatia, Disha and Becker, Mathias and Belyaev, Alexander and Benoit, Ferdinand and Blanke, Monika and Burzynski, Jackson and Butterworth, Jonathan M. and Cagnotta, Antimo and Calibbi, Lorenzo and Carpenter, Linda M. and Cid Vidal, Xabier and Copello, Emanuele and Corpe, Louie and D’Amico, Francesco and Deandrea, Aldo and Desai, Aman and Doglioni, Caterina and Dogra, Sunil M. and Garny, Mathias and Goodsell, Mark D. and Hassan, Sohaib and Harris, Philip Coleman and Harz, Julia and Ibarra, Alejandro and Iorio, Alberto Orso Maria and Kahlhoefer, Felix and Kar, Deepak and Khalil, Shaaban and Khoze, Valery and Ko, Pyungwon and Kraml, Sabine and Landsberg, Greg and Lessa, Andre and Lopez-Honorez, Laura and Mariotti, Alberto and Mitsou, Vasiliki A. and Mohan, Kirtimaan and Moon, Chang-Seong and Moreno Briceno, Alexander and Moreno Llacer, Maria and Munoz-Aillaud, Leandre and Murphy, Taylor and Ncube, Anele M. and Nzuza, Wandile and Prat, Clarisse and Rathmann, Lena and Sangweni, Thobani and Sengupta, Dipan and Shepherd, William and Sinha, Sukanya and Tait, Tim M. P. and Thamm, Andrea and Tytgat, Michel H. G. and Wang, Zirui and Yu, David and Yu, Shin-Shan},
      year={2025},
      month=sep,
      eprint         = "2504.10597",
      archivePrefix  = "arXiv",
      primaryClass   = "hep-ph" 
}

@article{HV-pythia,
   title={Visible effects of invisible Hidden Valley radiation},
   volume={09},
   % ISSN={1029-8479},
   % url={http://dx.doi.org/10.1007/JHEP09(2010)105},
   DOI={10.1007/jhep09(2010)105},
   pages={105},
   journal={JHEP},
   publisher={Springer Science and Business Media LLC},
   author={Carloni, Lisa and Sjöstrand, Torbjörn},
   year={2010},
   month=sep,
   eprint         = "1006.2911",
   archivePrefix  = "arXiv",
   primaryClass   = "hep-ph"  
}

@article{KalmanFilter,
   author  = {R. Frühwirth},
   title   = {Application of Kalman filtering to track and vertex fitting},
   journal = {Nucl. Instrum. Meth. A},
   year    = {1987},
   volume  = {262},
   pages   = {444--450},
   doi     = {10.1016/0168-9002(87)90887-4},
}

@article{Gross_Wilczek,
  title = {Ultraviolet Behavior of Non-Abelian Gauge Theories},
  author = {Gross, David J. and Wilczek, Frank},
  journal = {Phys. Rev. Lett.},
  volume = {30},
%   issue = {26},
  pages = {1343--1346},
  numpages = {0},
  year = {1973},
  month = {Jun},
  publisher = {American Physical Society},
  doi = {10.1103/PhysRevLett.30.1343},
%   url = {https://link.aps.org/doi/10.1103/PhysRevLett.30.1343}
}

@article{Politzer,
    title = {Reliable Perturbative Results for Strong Interactions?},
    author = {Politzer, H. David},
    journal = {Phys. Rev. Lett.},
    volume = {30},
   %  issue = {26},
    pages = {1346--1349},
    numpages = {0},
    year = {1973},
    month = {Jun},
    publisher = {American Physical Society},
    doi = {10.1103/PhysRevLett.30.1346},
   %  url = {https://link.aps.org/doi/10.1103/PhysRevLett.30.1346}
}

@article{Ball:2012cx,
      author         = "{NNPDF Collaboration} and Ball, Richard D. and others",
      title          = "{Parton distributions with LHC data}",
      journal        = "Nucl. Phys. B",
      volume         = "867",
      year           = "2013",
      pages          = "244",
      doi            = "10.1016/j.nuclphysb.2012.10.003",
      eprint         = "1207.1303",
      archivePrefix  = "arXiv",
      primaryClass   = "hep-ph",
      reportNumber   = "EDINBURGH-2012-08, IFUM-FT-997, FR-PHENO-2012-014,
                        RWTH-TTK-12-25, CERN-PH-TH-2012-037, SFB-CPP-12-47\,
                        --CERN-PH-TH-2012-037",
      SLACcitation   = "%%CITATION = ARXIV:1207.1303;%%"
}

@Article{Sjostrand:2007gs,
	author    = "Sj{\"o}strand, T. and Mrenna, S. and Skands, P.",
	title     = "{A brief introduction to PYTHIA 8.1}",
	journal   = "Comput. Phys. Commun.",
	volume    = "178",
	year      = "2008",
	pages     = "852-867",
	eprint    = "0710.3820",
	archivePrefix = "arXiv",
	primaryClass  =  "hep-ph",
	doi       = "10.1016/j.cpc.2008.01.036",
	SLACcitation  = "%%CITATION = 0710.3820;%%"
}

@article{Sjostrand:2014zea,
      author         = "Sj{\"o}strand, Torbj{\"o}rn and Ask, Stefan and Christiansen,
                        Jesper R. and Corke, Richard and Desai, Nishita and Ilten,
                        Philip and Mrenna, Stephen and Prestel, Stefan and
                        Rasmussen, Christine O. and Skands, Peter Z.",
      title          = "{An introduction to PYTHIA 8.2}",
      journal        = "Comput. Phys. Commun.",
      volume         = "191",
      year           = "2015",
      pages          = "159",
      doi            = "10.1016/j.cpc.2015.01.024",
      eprint         = "1410.3012",
      archivePrefix  = "arXiv",
      primaryClass   = "hep-ph",
      reportNumber   = "LU-TP-14-36, MCNET-14-22, CERN-PH-TH-2014-190,
                        FERMILAB-PUB-14-316-CD, DESY-14-178, SLAC-PUB-16122,
                        --FERMILAB-PUB-14-316-CD",
      SLACcitation   = "%%CITATION = ARXIV:1410.3012;%%"
}

@Article{Lange:2001uf,
      author         = "Lange, D. J.",
      title          = "{The EvtGen particle decay simulation package}",
      booktitle      = "{Proceedings, 7th International Conference on B physics
                        at hadron machines (BEAUTY 2000)}",
      journal        = "Nucl. Instrum. Meth. A",
      volume         = "462",
      year           = "2001",
      pages          = "152",
      doi            = "10.1016/S0168-9002(01)00089-4",
      SLACcitation   = "%%CITATION = NUIMA,A462,152;%%"
}

@Article{Nason:2004rx,
      author         = "Nason, Paolo",
      title          = "{A new method for combining NLO QCD with shower Monte Carlo algorithms}",
      journal        = "JHEP",
      volume         = "11",
      pages          = "040",
      doi            = "10.1088/1126-6708/2004/11/040",
      year           = "2004",
      eprint         = "hep-ph/0409146",
      archivePrefix  = "arXiv",
}

@Article{Frixione:2007vw,
      author         = "Frixione, Stefano and Nason, Paolo and Oleari, Carlo",
      title          = "{Matching NLO QCD computations with parton shower
                        simulations: the POWHEG method}",
      journal        = "JHEP",
      volume         = "11",
      pages          = "070",
      doi            = "10.1088/1126-6708/2007/11/070",
      year           = "2007",
      eprint         = "0709.2092",
      archivePrefix  = "arXiv",
      primaryClass   = "hep-ph",
}

@Article{Alioli:2010xd,
      author         = "Alioli, Simone and Nason, Paolo and Oleari, Carlo and Re,
                        Emanuele",
      title          = "{A general framework for implementing NLO calculations in
                        shower Monte Carlo programs: the POWHEG BOX}",
      journal        = "JHEP",
      volume         = "06",
      pages          = "043",
      doi            = "10.1007/JHEP06(2010)043",
      year           = "2010",
      eprint         = "1002.2581",
      archivePrefix  = "arXiv",
      primaryClass   = "hep-ph",
}

@article{Agostinelli:2002hh,
      author         = "Agostinelli, S. and others",
      title          = "{\textsc{Geant4} -- a simulation toolkit}",
      journal        = "Nucl. Instrum. Meth. A",
      volume         = "506",
      year           = "2003",
      pages          = "250",
      doi            = "10.1016/S0168-9002(03)01368-8",
      reportNumber   = "SLAC-PUB-9350, FERMILAB-PUB-03-339",
      SLACcitation   = "%%CITATION = NUIMA,A506,250;%%"
}

@Article{Read:2002hq,
    Author = {Read, Alexander L.},
    Journal = {J. Phys. G},
    Pages = {2693},
    doi = "10.1088/0954-3899/28/10/313",
    Title = "{Presentation of search results: the \(CL_s\) technique}",
    Volume = {28},
    Year = {2002}
}

@Booklet{TMVA,
      author         = "Hoecker, Andreas and others",
      title          = "{TMVA - Toolkit for Multivariate Data Analysis}",
      year           = "2009",
      eprint         = "physics/0703039",
      archivePrefix  = "arXiv",
      primaryClass   = "physics.data-an",
}

@Article{PERF-2007-01,
    author         = "{ATLAS Collaboration}",
    title          = "{The ATLAS Experiment at the CERN Large Hadron Collider}",
    journal        = "JINST",
    volume         = "3",
    year           = "2008",
    pages          = "S08003",
    doi            = "10.1088/1748-0221/3/08/S08003",
    primaryClass   = "hep-ex",
}

@Article{SOFT-2010-01,
    author         = "{ATLAS Collaboration}",
    title          = "{The ATLAS Simulation Infrastructure}",
    journal        = "Eur. Phys. J. C",
    volume         = "70",
    year           = "2010",
    pages          = "823",
    doi            = "10.1140/epjc/s10052-010-1429-9",
    eprint         = "1005.4568",
    archivePrefix  = "arXiv",
    primaryClass   = "physics.ins-det",
}

@Article{TRIG-2012-01,
    author         = "{ATLAS Collaboration}",
    title          = "{The performance of the jet trigger for the ATLAS detector during 2011 data taking}",
    journal        = "Eur. Phys. J. C",
    volume         = "76",
    year           = "2016",
    pages          = "526",
    doi            = "10.1140/epjc/s10052-016-4325-0",
    reportNumber   = "CERN-EP-2016-031",
    eprint         = "1606.07759",
    archivePrefix  = "arXiv",
    primaryClass   = "hep-ex",
}

@Article{EXOT-2013-12,
    author         = "{ATLAS Collaboration}",
    title          = "{Search for long-lived, weakly interacting particles that decay to displaced hadronic jets in proton--proton collisions at \(\sqrt{s} = 8\,\text{TeV}\) with the ATLAS detector}",
    journal        = "Phys. Rev. D",
    volume         = "92",
    year           = "2015",
    pages          = "012010",
    doi            = "10.1103/PhysRevD.92.012010",
    reportNumber   = "CERN-PH-EP-2015-071",
    eprint         = "1504.03634",
    archivePrefix  = "arXiv",
    primaryClass   = "hep-ex",
}

@Article{PERF-2014-07,
    author         = "{ATLAS Collaboration}",
    title          = "{Topological cell clustering in the ATLAS calorimeters and its performance in LHC Run~1}",
    journal        = "Eur. Phys. J. C",
    volume         = "77",
    year           = "2017",
    pages          = "490",
    doi            = "10.1140/epjc/s10052-017-5004-5",
    reportNumber   = "CERN-PH-EP-2015-304",
    eprint         = "1603.02934",
    archivePrefix  = "arXiv",
    primaryClass   = "hep-ex",
}

@Article{PERF-2015-07,
    author         = "{ATLAS Collaboration}",
    title          = "{Study of the material of the ATLAS inner detector for Run~2 of the LHC}",
    journal        = "JINST",
    volume         = "12",
    year           = "2017",
    pages          = "P12009",
    doi            = "10.1088/1748-0221/12/12/P12009",
    reportNumber   = "CERN-EP-2017-081",
    eprint         = "1707.02826",
    archivePrefix  = "arXiv",
    primaryClass   = "hep-ex",
}

@Article{STDM-2015-05,
    author         = "{ATLAS Collaboration}",
    title          = "{Measurement of the Inelastic Proton--Proton Cross Section at \(\sqrt{s} = 13\,\text{TeV}\) with the ATLAS Detector at the LHC}",
    journal        = "Phys. Rev. Lett.",
    volume         = "117",
    year           = "2016",
    pages          = "182002",
    doi            = "10.1103/PhysRevLett.117.182002",
    reportNumber   = "CERN-EP-2016-140",
    eprint         = "1606.02625",
    archivePrefix  = "arXiv",
    primaryClass   = "hep-ex",
}

@Article{PERF-2016-04,
    author         = "{ATLAS Collaboration}",
    title          = "{Jet energy scale measurements and their systematic uncertainties in proton--proton collisions at \(\sqrt{s} = 13\,\text{TeV}\) with the ATLAS detector}",
    journal        = "Phys. Rev. D",
    volume         = "96",
    year           = "2017",
    pages          = "072002",
    doi            = "10.1103/PhysRevD.96.072002",
    reportNumber   = "CERN-EP-2017-038",
    eprint         = "1703.09665",
    archivePrefix  = "arXiv",
    primaryClass   = "hep-ex",
}

@Article{TRIG-2016-01,
    author         = "{ATLAS Collaboration}",
    title          = "{Performance of the ATLAS trigger system in 2015}",
    journal        = "Eur. Phys. J. C",
    volume         = "77",
    year           = "2017",
    pages          = "317",
    doi            = "10.1140/epjc/s10052-017-4852-3",
    reportNumber   = "CERN-EP-2016-241",
    eprint         = "1611.09661",
    archivePrefix  = "arXiv",
    primaryClass   = "hep-ex",
}

@Article{DAPR-2018-01,
    author         = "{ATLAS Collaboration}",
    title          = "{ATLAS data quality operations and performance for 2015--2018 data-taking}",
    journal        = "JINST",
    volume         = "15",
    year           = "2020",
    pages          = "P04003",
    doi            = "10.1088/1748-0221/15/04/P04003",
    reportNumber   = "CERN-EP-2019-207",
    eprint         = "1911.04632",
    archivePrefix  = "arXiv",
    primaryClass   = "physics.ins-det",
}

@Article{EXOT-2018-20,
    author         = "{ATLAS Collaboration}",
    title          = "{Search for emerging jets in \(pp\) collisions at \(\sqrt{s} = 13\,\text{TeV}\) with the ATLAS experiment}",
    year           = "2025",
    reportNumber   = "CERN-EP-2025-225",
    eprint         = "2510.12347",
    archivePrefix  = "arXiv",
    primaryClass   = "hep-ex",
}

@Article{EXOT-2018-57,
    author         = "{ATLAS Collaboration}",
    title          = "{Search for exotic decays of the Higgs boson into long-lived particles in \(pp\) collisions at \(\sqrt{s} = 13\,\text{TeV}\) using displaced vertices in the ATLAS inner detector}",
    journal        = "JHEP",
    volume         = "11",
    year           = "2021",
    pages          = "229",
    doi            = "10.1007/JHEP11(2021)229",
    reportNumber   = "CERN-EP-2021-106",
    eprint         = "2107.06092",
    archivePrefix  = "arXiv",
    primaryClass   = "hep-ex",
}

@Article{HDBS-2018-45,
    author         = "{ATLAS Collaboration}",
    title          = "{Search for resonant production of dark quarks in the dijet final state with the ATLAS detector}",
    journal        = "JHEP",
    volume         = "02",
    year           = "2024",
    pages          = "128",
    doi            = "10.1007/JHEP02(2024)128",
    reportNumber   = "CERN-EP-2023-217",
    eprint         = "2311.03944",
    archivePrefix  = "arXiv",
    primaryClass   = "hep-ex",
}

@Article{JETM-2018-02,
    author         = "{ATLAS Collaboration}",
    title          = "{In situ calibration of large-radius jet energy and mass in \(13\,\text{TeV}\) proton--proton collisions with the ATLAS detector}",
    journal        = "Eur. Phys. J. C",
    volume         = "79",
    year           = "2019",
    pages          = "135",
    doi            = "10.1140/epjc/s10052-019-6632-8",
    reportNumber   = "CERN-EP-2018-191",
    eprint         = "1807.09477",
    archivePrefix  = "arXiv",
    primaryClass   = "hep-ex",
}

@Article{JETM-2018-05,
    author         = "{ATLAS Collaboration}",
    title          = "{Jet energy scale and resolution measured in proton--proton collisions at \(\sqrt{s} = 13\,\text{TeV}\) with the ATLAS detector}",
    journal        = "Eur. Phys. J. C",
    volume         = "81",
    year           = "2021",
    pages          = "689",
    doi            = "10.1140/epjc/s10052-021-09402-3",
    reportNumber   = "CERN-EP-2020-083",
    eprint         = "2007.02645",
    archivePrefix  = "arXiv",
    primaryClass   = "hep-ex",
}

@Article{PIX-2018-001,
    author         = "Abbott, B. and others",
    title          = "{Production and integration of the ATLAS Insertable B-Layer}",
    journal        = "JINST",
    volume         = "13",
    year           = "2018",
    pages          = "T05008",
    doi            = "10.1088/1748-0221/13/05/T05008",
    eprint         = "1803.00844",
    archivePrefix  = "arXiv",
    primaryClass   = "physics.ins-det",
}

@Article{FTAG-2019-07,
    author         = "{ATLAS Collaboration}",
    title          = "{ATLAS flavour-tagging algorithms for the LHC Run~2 \(pp\) collision dataset}",
    journal        = "Eur. Phys. J. C",
    volume         = "83",
    year           = "2023",
    pages          = "681",
    doi            = "10.1140/epjc/s10052-023-11699-1",
    reportNumber   = "CERN-EP-2022-226",
    eprint         = "2211.16345",
    archivePrefix  = "arXiv",
    primaryClass   = "physics.data-an",
}

@Article{STDM-2019-02,
    author         = "{ATLAS Collaboration}",
    title          = "{Measurement of hadronic event shapes in high-\(p_{\text{T}}\) multijet final states at \( \sqrt{s} = 13\,\text{TeV}\) with the ATLAS detector}",
    journal        = "JHEP",
    volume         = "01",
    year           = "2021",
    pages          = "188",
    doi            = "10.1007/JHEP01(2021)188",
    reportNumber   = "CERN-EP-2020-077",
    eprint         = "2007.12600",
    archivePrefix  = "arXiv",
    primaryClass   = "hep-ex",
    related        = "STDM-2019-02-err",
    relatedstring  = "Erratum:",
}

@Article{TDAQ-2019-01,
    author         = "{ATLAS Collaboration}",
    title          = "{Performance of the upgraded PreProcessor of the ATLAS Level-1 Calorimeter Trigger}",
    journal        = "JINST",
    volume         = "15",
    year           = "2020",
    pages          = "P11016",
    doi            = "10.1088/1748-0221/15/11/P11016",
    reportNumber   = "CERN-EP-2020-042",
    eprint         = "2005.04179",
    archivePrefix  = "arXiv",
    primaryClass   = "physics.ins-det",
}

@Article{TRIG-2019-04,
    author         = "{ATLAS Collaboration}",
    title          = "{Operation of the ATLAS trigger system in Run~2}",
    journal        = "JINST",
    volume         = "15",
    year           = "2020",
    pages          = "P10004",
    doi            = "10.1088/1748-0221/15/10/P10004",
    reportNumber   = "CERN-EP-2020-109",
    eprint         = "2007.12539",
    archivePrefix  = "arXiv",
    primaryClass   = "physics.ins-det",
}

@Article{DAPR-2021-01,
    author         = "{ATLAS Collaboration}",
    title          = "{Luminosity determination in \(pp\) collisions at \(\sqrt{s} = 13\,\text{TeV}\) using the ATLAS detector at the LHC}",
    journal        = "Eur. Phys. J. C",
    volume         = "83",
    year           = "2023",
    pages          = "982",
    doi            = "10.1140/epjc/s10052-023-11747-w",
    reportNumber   = "CERN-EP-2022-281",
    eprint         = "2212.09379",
    archivePrefix  = "arXiv",
    primaryClass   = "hep-ex",
}

@Article{EXOT-2021-19,
    author         = "{ATLAS Collaboration}",
    title          = "{Search for new physics in final states with semivisible jets or anomalous signatures using the ATLAS detector}",
    journal        = "Phys. Rev. D",
    volume         = "112",
    year           = "2025",
    pages          = "012021",
    doi            = "10.1103/44zp-mh1q",
    reportNumber   = "CERN-EP-2025-101",
    eprint         = "2505.01634",
    archivePrefix  = "arXiv",
    primaryClass   = "hep-ex",
}

@Article{EXOT-2022-31,
    author         = "{ATLAS Collaboration}",
    title          = "{Search for emerging jets in \(pp\) collisions at \(\sqrt{s} = 13.6\,\text{TeV}\) with the ATLAS experiment}",
    journal        = "Rept. Prog. Phys.",
    volume         = "88",
    year           = "2025",
    pages          = "097801",
    doi            = "10.1088/1361-6633/adfe17",
    reportNumber   = "CERN-EP-2025-099",
    eprint         = "2505.02429",
    archivePrefix  = "arXiv",
    primaryClass   = "hep-ex",
}

@Article{EXOT-2022-37,
    author         = "{ATLAS Collaboration}",
    title          = "{Search for non-resonant production of semi-visible jets using Run~2 data in ATLAS}",
    journal        = "Phys. Lett. B",
    volume         = "848",
    year           = "2024",
    pages          = "138324",
    doi            = "10.1016/j.physletb.2023.138324",
    reportNumber   = "CERN-EP-2023-084",
    eprint         = "2305.18037",
    archivePrefix  = "arXiv",
    primaryClass   = "hep-ex",
}

@Article{EXOT-2023-14,
    author         = "{ATLAS Collaboration}",
    title          = "{Exploration at the high-energy frontier: ATLAS Run~2 searches investigating the exotic jungle beyond the Standard Model}",
    journal        = "Phys. Rept.",
    volume         = "1116",
    year           = "2025",
    pages          = "301--385",
    doi            = "10.1016/j.physrep.2024.10.001",
    reportNumber   = "CERN-EP-2024-075",
    eprint         = "2403.09292",
    archivePrefix  = "arXiv",
    primaryClass   = "hep-ex",
}

@Booklet{ATL-SOFT-PUB-2025-001,
    author         = "{ATLAS Collaboration}",
    title          = "{ATLAS Computing Acknowledgements}",
    howpublished   = "{ATL-SOFT-PUB-2025-001}",
    url            = "https://cds.cern.ch/record/2922210",
    year           = "2025",
}

@Report{ATLAS-TDR-19,
    author         = "{ATLAS Collaboration}",
    title          = "{ATLAS Insertable B-Layer: Technical Design Report}",
    type           = "ATLAS-TDR-19; CERN-LHCC-2010-013",
    year           = "2010",
    url            = "https://cds.cern.ch/record/1291633",
    related        = "ATLAS-TDR-19-addm",
    relatedstring  = "Addendum:",
}

@Article{LUCID2,
    author         = "Avoni, G. and others",
    title          = "{The new LUCID-2 detector for luminosity measurement and monitoring in ATLAS}",
    journal        = "JINST",
    volume         = "13",
    year           = "2018",
    pages          = "P07017",
    doi            = "10.1088/1748-0221/13/07/P07017",
}

@Article{CMS-EXO-18-001,
    author         = "{CMS Collaboration}",
    title          = "{Search for new particles decaying to a jet and an emerging jet}",
    journal        = "JHEP",
    volume         = "02",
    year           = "2019",
    pages          = "179",
    doi            = "10.1007/JHEP02(2019)179",
    reportNumber   = "CERN-EP-2018-255",
    eprint         = "1810.10069",
    archivePrefix  = "arXiv",
    primaryClass   = "hep-ex",
}

@Article{CMS-EXO-19-020,
    author         = "{CMS Collaboration}",
    title          = "{Search for resonant production of strongly coupled dark matter in proton--proton collisions at \(13\,\text{TeV}\)}",
    journal        = "JHEP",
    volume         = "06",
    year           = "2022",
    pages          = "156",
    doi            = "10.1007/JHEP06(2022)156",
    reportNumber   = "CERN-EP-2021-252",
    eprint         = "2112.11125",
    archivePrefix  = "arXiv",
    primaryClass   = "hep-ex",
}

@Article{CMS-EXO-22-015,
    author         = "{CMS Collaboration}",
    title          = "{Search for dark QCD with emerging jets in proton--proton collisions at \(\sqrt{s} = 13\,\text{TeV}\)}",
    journal        = "JHEP",
    volume         = "07",
    year           = "2024",
    pages          = "142",
    doi            = "10.1007/JHEP07(2024)142",
    reportNumber   = "CERN-EP-2024-049",
    eprint         = "2403.01556",
    archivePrefix  = "arXiv",
    primaryClass   = "hep-ex",
}

@Article{CMS-EXO-23-005,
    author         = "{CMS Collaboration}",
    title          = "{Dark sector searches with the CMS experiment}",
    journal        = "Phys. Rept.",
    volume         = "1115",
    year           = "2025",
    pages          = "448--569",
    doi            = "10.1016/j.physrep.2024.09.013",
    reportNumber   = "CERN-EP-2024-106",
    eprint         = "2405.13778",
    archivePrefix  = "arXiv",
    primaryClass   = "hep-ex",
}

@Booklet{ATL-PHYS-PUB-2014-021,
    author         = "{ATLAS Collaboration}",
    title          = "{ATLAS Pythia~8 tunes to \(7\,\text{TeV}\) data}",
    howpublished   = "{ATL-PHYS-PUB-2014-021}",
    url            = "https://cds.cern.ch/record/1966419",
    year           = "2014",
}

@Booklet{ATL-PHYS-PUB-2015-026,
    author         = "{ATLAS Collaboration}",
    title          = "{Vertex Reconstruction Performance of the ATLAS Detector at \(\sqrt{s} = 13\,\text{TeV}\)}",
    howpublished   = "{ATL-PHYS-PUB-2015-026}",
    url            = "https://cds.cern.ch/record/2037717",
    year           = "2015",
}

@Booklet{ATL-PHYS-PUB-2015-051,
    author         = "{ATLAS Collaboration}",
    title          = "{Early Inner Detector Tracking Performance in the 2015 Data at \(\sqrt{s} = 13\,\text{TeV}\)}",
    howpublished   = "{ATL-PHYS-PUB-2015-051}",
    url            = "https://cds.cern.ch/record/2110140",
    year           = "2015",
}

@Booklet{ATL-PHYS-PUB-2016-017,
    author         = "{ATLAS Collaboration}",
    title          = "{The Pythia~8 A3 tune description of ATLAS minimum bias and inelastic measurements incorporating the Donnachie--Landshoff diffractive model}",
    howpublished   = "{ATL-PHYS-PUB-2016-017}",
    url            = "https://cds.cern.ch/record/2206965",
    year           = "2016",
}

@Booklet{ATL-PHYS-PUB-2017-014,
    author         = "{ATLAS Collaboration}",
    title          = "{Performance of the reconstruction of large impact parameter tracks in the inner detector of ATLAS}",
    howpublished   = "{ATL-PHYS-PUB-2017-014}",
    url            = "https://cds.cern.ch/record/2275635",
    year           = "2017",
}

@Booklet{ATL-PHYS-PUB-2019-013,
    author         = "{ATLAS Collaboration}",
    title          = "{Performance of vertex reconstruction algorithms for detection of new long-lived particle decays within the ATLAS inner detector}",
    howpublished   = "{ATL-PHYS-PUB-2019-013}",
    url            = "https://cds.cern.ch/record/2669425",
    year           = "2019",
}

@Booklet{ATL-PHYS-PUB-2019-017,
    author         = "{ATLAS Collaboration}",
    title          = "{Multijet simulation for \(13\,\text{TeV}\) ATLAS Analyses}",
    howpublished   = "{ATL-PHYS-PUB-2019-017}",
    url            = "https://cds.cern.ch/record/2672252",
    year           = "2019",
}

@Booklet{ATL-PHYS-PUB-2020-023,
    author         = "{ATLAS Collaboration}",
    title          = "{Study of top-quark pair modelling and uncertainties using ATLAS measurements at \(\sqrt{s} = 13\,\text{TeV}\)}",
    howpublished   = "{ATL-PHYS-PUB-2020-023}",
    url            = "https://cds.cern.ch/record/2730443",
    year           = "2020",
}
\clearpage
 
\begin{flushleft}
\hypersetup{urlcolor=black}
{\Large The ATLAS Collaboration}

\bigskip

\AtlasOrcid[0000-0002-6665-4934]{G.~Aad}$^\textrm{\scriptsize 103}$,
\AtlasOrcid[0000-0001-7616-1554]{E.~Aakvaag}$^\textrm{\scriptsize 17}$,
\AtlasOrcid[0000-0002-5888-2734]{B.~Abbott}$^\textrm{\scriptsize 122}$,
\AtlasOrcid[0000-0002-0287-5869]{S.~Abdelhameed}$^\textrm{\scriptsize 118a}$,
\AtlasOrcid[0000-0002-1002-1652]{K.~Abeling}$^\textrm{\scriptsize 55}$,
\AtlasOrcid[0000-0001-5763-2760]{N.J.~Abicht}$^\textrm{\scriptsize 49}$,
\AtlasOrcid[0000-0002-8496-9294]{S.H.~Abidi}$^\textrm{\scriptsize 30}$,
\AtlasOrcid[0009-0003-6578-220X]{M.~Aboelela}$^\textrm{\scriptsize 45}$,
\AtlasOrcid[0000-0002-9987-2292]{A.~Aboulhorma}$^\textrm{\scriptsize 36e}$,
\AtlasOrcid[0000-0001-5329-6640]{H.~Abramowicz}$^\textrm{\scriptsize 156}$
% ...
\bigskip
\\

% ...
$^{17}$Department for Physics and Technology, University of Bergen, Bergen; Norway.\\
% ...
$^{30}$Physics Department, Brookhaven National Laboratory, Upton NY; United States of America.\\
% ...
$^{36}$$^{(a)}$Facult\'e des Sciences Ain Chock, Universit\'e Hassan II de Casablanca;$^{(b)}$Facult\'{e} des Sciences, Universit\'{e} Ibn-Tofail, K\'{e}nitra;$^{(c)}$Facult\'e des Sciences Semlalia, Universit\'e Cadi Ayyad, LPHEA-Marrakech;$^{(d)}$LPMR, Facult\'e des Sciences, Universit\'e Mohamed Premier, Oujda;$^{(e)}$Facult\'e des sciences, Universit\'e Mohammed V, Rabat;$^{(f)}$Institute of Applied Physics, Mohammed VI Polytechnic University, Ben Guerir; Morocco.\\
% ...
$^{45}$Physics Department, Southern Methodist University, Dallas TX; United States of America.\\
% ...
$^{49}$Fakult\"{a}t Physik, Technische Universit{\"a}t Dortmund, Dortmund; Germany.\\
% ...
$^{55}$II. Physikalisches Institut, Georg-August-Universit\"{a}t G\"ottingen, G\"ottingen; Germany.\\
% ...
$^{103}$CPPM, Aix-Marseille Universit\'e, CNRS/IN2P3, Marseille; France.\\
% ...
$^{118}$$^{(a)}$New York University Abu Dhabi, Abu Dhabi;$^{(b)}$United Arab Emirates University, Al Ain; United Arab Emirates.\\
% ...
$^{122}$Homer L. Dodge Department of Physics and Astronomy, University of Oklahoma, Norman OK; United States of America.\\
% ...
$^{156}$Raymond and Beverly Sackler School of Physics and Astronomy, Tel Aviv University, Tel Aviv; Israel.\\
% ...
$^{a}$ Also at Affiliated with an institute formerly covered by a cooperation agreement with CERN.\\
$^{b}$ Also at An-Najah National University, Nablus; Palestine.\\
$^{c}$ Also at Borough of Manhattan Community College, City University of New York, New York NY; United States of America.\\
$^{d}$ Also at Center for Interdisciplinary Research and Innovation (CIRI-AUTH), Thessaloniki; Greece.\\
$^{e}$ Also at Centre of Physics of the Universities of Minho and Porto (CF-UM-UP); Portugal.\\
$^{f}$ Also at CERN, Geneva; Switzerland.\\
% ...

\end{flushleft}


%

%
%

%

%
%
%
%

\end{document}